\documentclass[aps,pre,reprint,amsmath,amssymb,longbibliography,lengthcheck,showpacs,superscriptaddress,floatfix]{revtex4-2}
\usepackage{enumerate}
\usepackage{color}
\usepackage{physics}
\usepackage{graphicx}
\begin{document}
%
\title{Structural fluctuations in active glasses}
%
\author{Masaki Yoshida}
\email{yoshida-masaki55@g.ecc.u-tokyo.ac.jp}
\affiliation{Graduate School of Arts and Sciences, The University of Tokyo, Tokyo 153-8902, Japan}
\author{Hideyuki Mizuno}
\email{hideyuki.mizuno@phys.c.u-tokyo.ac.jp}
\affiliation{Graduate School of Arts and Sciences, The University of Tokyo, Tokyo 153-8902, Japan}
\author{Atsushi Ikeda}
\email{atsushi.ikeda@phys.c.u-tokyo.ac.jp}
\affiliation{Graduate School of Arts and Sciences, The University of Tokyo, Tokyo 153-8902, Japan}
\affiliation{Research Center for Complex Systems Biology, Universal Biology Institute, The University of Tokyo, Tokyo 153-8902, Japan}
%
\date{\today}
%
\begin{abstract}
The glassy dynamics of dense active matter have recently become a topic of interest due to their importance in biological processes such as wound healing and tissue development.
However, while the liquid-state properties of dense active matter have been studied in relation to the glass transition of active matter, the solid-state properties of active glasses have yet to be understood.
In this work, we study the structural fluctuations in the active glasses composed of self-propelled particles. 
We develop a formalism to describe the solid-state properties of active glasses in the harmonic approximation limit and use it to analyze the displacement fields in the active glasses. 
Our findings reveal that the dynamics of high-frequency normal modes become quasi-static with respect to the active forces, and consequently, excitations of these modes are significantly suppressed.
This leads to a violation of the equipartition law, suppression of particle displacements, and the apparent collective motion of active glasses. 
Overall, our results provide a fundamental understanding of the solid-state properties of active glasses.
\end{abstract}
%
\maketitle
%

\section{Introduction}
\textit{Active matter} refers to materials and systems that are capable of converting energy into autonomous motion~\cite{Marchetti2013}.
Active matter can be found abundantly in our natural world, ranging from macroscopic organisms like flying birds and swimming fish to microscopic systems like cytoplasm and bacteria.
In these systems, each component or particle is driven by non-thermal, active forces, which cause a number of novel non-equilibrium phenomena~\cite{Vicsek_1995,Cates2015,Alert2022,Ramaswamy_2003}. 

When the active matter is in sufficiently dense states, it often exhibits glassy dynamics~\cite{Angelini_2011,Berthier2019,Janssen_2019}.
This phenomenon has received considerable attention in biological physics, especially in studies of the collective motion of confluent epithelial cells that regulate wound healing, cancer metastasis, and other processes~\cite{Angelini_2011,Zhou_2009,Nnetu_2012,Schoetz_2013,Haeger2014,Garcia_2015,Park_2016,Vishwakarma_2020}, the diffusion and viscoelasticity of proteins in bacterial cytoplasm~\cite{Parry2014, Nishizawa2017}, and the dynamics and rheology of dense bacterial suspensions~\cite{Lama_2023, Sugino2024}.
Correspondingly, many numerical and theoretical models of dense active matter have been proposed and studied~\cite{Henkes_2011,Bi2016,Oyama2019}. 
Among them, the simplest is presumably the self-propelled particles. 
Previous works have established that this model shows canonical glassy dynamics when the density is high enough or the active force is weak enough.
In fact, the abrupt slowdown of the microscopic dynamics occers, the dynamic heterogeneity emerges, and the system eventually vitrifies into a glass state called the \textit{active glass}.
However, several non-canonical dynamical behaviors have also been observed, such as a change in the glass transition density, violation of the fluctuation-dissipation theorem, and the emergence of novel collective dynamics~\cite{Henkes_2011,Ni2013,Berthier_2014,Flenner_2016,Nandi_2017,Nandi_2018,Mandal_2020,Klongvessa_2022,Paul_2023,Keta_2023,Paoluzzi2024}.
Extensions of equilibrium theories of the glass transition have been proposed to understand these behaviors~\cite{Berthier_2013,Nandi_2018,Paul_2023,Nandi_2017,Liluashvili_2017}, but they are still the subject of intense debate.

As mentioned above, previous studies have focused on the properties of dense active matter in their liquid state, while little is known about the properties of active glasses in their solid state.
The present study aims to shed light on the structural fluctuations in active glasses.
In passive glasses, particles are driven solely by thermal agitations, and their structural fluctuations are determined by their vibrational normal modes.
Glasses have a large number of non-phonon normal modes, which differ significantly from crystalline solids where the normal modes are longitudinal and transverse phonons, and their vibrational density of states (vDOS) follows the Debye law $g(\omega) \propto \omega^{d-1}$~(where $d$ is the spatial dimension).
As a result, there exist excess vibrational states over the Debye law, which is observed by a large peak in the reduced vDOS $g(\omega)/\omega^{d-1}$, referred to as the boson peak~\cite{Buchenau_1984,Yamamuro_1996,Mori_2020}.
Additionally, at much lower frequencies, the vibrational normal modes become a mixture of phonon vibrations and quasi-localized vibrations~\cite{Lerner_2016,Mizuno_2017,Shimada_2018,Wang_2019}.
Within the harmonic description, structural fluctuations in passive glasses can be described in terms of the thermal excitations of these characteristic normal modes, all of which follow the equipartition law. 
Even in anharmonic dynamics at finite temperatures~\cite{Mizuno_2020}, the quasi-localized vibrations can be used to indicate structural rearrangements~\cite{Widmer_2008}. 
Therefore, to understand the structural fluctuations in active glasses, it is essential to examine how the active forces excite the characteristic normal modes in glasses.

In the present work, we study active glasses by employing three different models: Active Ornstein Uhlenbeck Particles~(AOUPs)~\cite{Martin_2021}, Active Brownian Particles~(ABPs)~\cite{Fily2012}, and Run and Tumble Particles~(RTPs)~\cite{Solon_2015}. 
To characterize structural fluctuations, we consider four observables: average excitation energy of the normal modes, mean-squared displacements~(MSD), structure factor of longitudinal and transverse displacement fields, and non-Gaussian parameters~(NGP).
Firstly, we demonstrate that the former three observables can be described in terms of the two-time correlations of active forces and the normal modes of glasses.
Next, we formulate the two-time correlations of the active forces to obtain general formulas of the three observables.
Additionally, we prove that AOUPs, ABPs, and RTPs have the same two-time correlation when the active forces are properly reparameterized.
We then apply the obtained formulas to calculate the observables for a typical configuration of active glasses.
Our central observation is that active forces behave like random noise for low-frequency normal modes, while they behave like constant forces for high-frequency normal modes.
This results in thermal excitations of low-frequency modes on one hand and quasi-static dynamics of high-frequency modes on the other, both of which determine nature of the structural fluctuations in active glasses.
Finally, we study the NGP and find that fluctuations are mostly Gaussian, except for normal modes at the high-frequency edge.
It should be noted that a previous study conducted by Henke \textit{et al.}~\cite{Henkes_2020} investigated the velocity fields in confluent cells, both experimentally and theoretically.
In their study, they modeled the cells using the ABPs, and demonstrated that the excitations of the normal modes in the model were able to accurately describe the velocity fields of cells as measured by experiments. 
The present work provides a more thorough analysis of the displacement fields, which complements and generalizes the work done by Henke \textit{et al.}~\cite{Henkes_2020}.

\section{Setting and observables}
The present work examines the fluctuation dynamics of particles in active glass models.
This section introduces our settings and observables.

\subsection{System}
We are considering a system of $N$ particles enclosed within a two-dimensional box ($d=2$) under periodic boundary conditions.
The configuration of the particles is represented by a $2N$-dimensional vector, $\vb*{r} = (\vec{r}_1, \vec{r}_2, \cdots, \vec{r}_N)$, where $\vec{r}_i$ describes the position of the particle $i$~($i=1,2,\cdots,N$).
The total potential energy of the system is denoted by ${U}(\vb*{r})$.
We study a mechanically stable configuration, also known as an inherent structure, $\vb*{r}_{0}$, which satisfies
\begin{equation}
\left. \frac{\partial {U}}{\partial \vb*{r}} \right|_{\vb*{r} = \vb*{r}_0} = \vb*{0}. 
\end{equation}
Our objective is to explore the dynamics of the active particles as they fluctuate around $\vb*{r}_{0}$.

\subsection{Equation of motion}
To mimic the active glasses, we adopt the following overdamped equation of motion:
\begin{equation}
\gamma \frac{d\vb*{r}}{dt} = - \pdv{U(\vb*{r})}{\vb*{r}} + \vb*{f}, \label{eqom1}
\end{equation}
where $t$ is time, and $\gamma$ represents the damping coefficient.
The $2N$-dimensional vector $\vb*{f} = (\vec{f}_1, \vec{f}_2, \cdots, \vec{f}_N)$ expresses the active forces.

When the active forces are sufficiently weak compared to the interparticle forces, the displacements of the particles, $\vb*{u} = \vb*{r} - \vb*{r}_{0}$, are small enough that the harmonic approximation limit is valid.
In the present work, we limit our discussion to this case. 
We first define a $2N\times 2N$ dynamical matrix, denoted as $\mathcal{M}$, as
\begin{equation} \label{dmat}
\left. \mathcal{M} = \frac{1}{\gamma} \pdv{U}{\vb*{r}}{\vb*{r}^\dag} \right|_{\vb*{r}=\vb*{r}_0}.
\end{equation}
where $\dag$ denotes conjugate operation.
Using the dynamical matrix, the potential energy can be written as
\begin{equation}
U = U_0 + \frac{\gamma}{2} \vb*{u}^\dag \mathcal{M} \vb*{u},
\end{equation}
where $U_0={U}(\vb*{r}=\vb*{r}_0)$ is constant. 
Then, the equation of motion, Eq.~(\ref{eqom1}), is simplifed into 
\begin{equation}
\gamma \frac{d\vb*{u}}{dt} = - \gamma \mathcal{M} \vb*{u} + \vb*{f}. \label{eqom2}
\end{equation}
Because we assumed the mechanical stability of the configuration $\vb*{r}_0$, the dynamical matrix $\mathcal{M}$ is positive semi-definite and has two translational zero modes. 
By removing these zero modes, we can focus on the fluctuation dynamics around $\vb*{u}=\vb*{0}$.
In this study, we solve Eq.~(\ref{eqom2}) to investigate the fluctuation dynamics in the steady state.
We consider three different active forces, each corresponding to a different active glass model.

\subsubsection{Active Ornstein Uhlenbeck Particles}
In the AOUPs~\cite{Martin_2021}, the active force follows
\begin{equation}
\tau \dv{\vb*{f}}{t} = - \vb*{f} + \sqrt{2 \gamma D} \vb*{\eta}, \label{eq:aoup}
\end{equation}
where $\tau$ expresses the persistent time of the active forces, and $D$ is the energy scale controlling the amplitude of the active forces. 
$\vb*{\eta}$ is the Gaussian white noise which statisfies $\langle \vb*{\eta}(t) \rangle = 0$ and $\langle \vb*{\eta}(t) \vb*{\eta}(t') \rangle = I \delta(t-t')$, where $\langle \rangle$ is ensemble average, $I$ is a $2N\times 2N$ unit tensor, and $\delta(t)$ is Dirac's delta function. 
From Eq.~(\ref{eq:aoup}), we can find
\begin{equation}
\langle \vb*{f}(t) \vb*{f}(t') \rangle = I \frac{\gamma D}{\tau} \exp \left( -\frac{|t-t'|}{\tau} \right).
\label{noise_correlation}
\end{equation}
In the limit of $\tau \to 0$, this active force reduces to the thermal noise, with the temperature $D$. 
Hereafter, we refer to the active glass model with this active force as the AOUP glass. 

\subsubsection{Active Brownian Particles}
In the ABPs~\cite{Fily2012}, the particle tries to maintain a constant speed while their direction changes constantly.
The active force acting on ABPs is described by 
\begin{equation}
\vb*{f} = \gamma v_0 (\cos{\theta_1}, \sin{\theta_1}, \cdots , \cos{\theta_N}, \sin{\theta_N}), \label{eq:abp}
\end{equation}
where $v_0$ is the constant self-propelled velocity in the non-interacting case. 
Notice $\vec{f}_i = (\gamma v_0 \cos{\theta_i}, \gamma v_0 \sin{\theta_i})$. 
The angle vector $\vb*{\theta} = (\theta_1, \theta_2, \cdots, \theta_N)$ follows the Wiener process as
\begin{equation} \label{eq.abp_theta}
\dv{\vb*{\theta}}{t} =  \sqrt{2D_R} \vb*{\eta},
\end{equation}
where $D_R$ is the rotational diffusion coefficient, and $\vb*{\eta}$ is the Gaussian white noise which is the same as in Eq.~(\ref{eq:aoup}). 
Hereafter, we refer to the active glass model with this active force as the ABP glass. 

\subsubsection{Run and Tumble Particles}
In the RTPs~\cite{Solon_2015}, the particle attempts to move at a constant speed, similar to the ABPs.
However, their directions change abruptly and randomly.
The force vector, $\vb*{f}$, is described by Eq.~(\ref{eq:abp}).
Meanwhile, each angle $\theta_i$ changes randomly in the range of $0\le \theta_i < 2\pi$ with a probability per unit time $P_\text{\rm tumble} = 1-e^{-R_{\rm tumble}}$, where $R_{\rm tumble}$ is a constant rate.
Hereafter, we refer to the active glass model with this active force as the RTP glass. 

\subsection{Normal modes}
Here we introduce the eigenmodes of the dynamical matrix $\mathcal{M}$ to solve Eq.~(\ref{eqom2}). 
We define the eigenvalues and eigenvectors by  
\begin{equation} \label{eq:eigenproblem}
\mathcal{M}\vb*{e}_l = \lambda_l\vb*{e}_l \qquad (l=1,2, \cdots, 2N-2), 
\end{equation}
where $\lambda_l$ and $\vb*{e}_l = \lparen \vec{e}_{l,1}, \vec{e}_{l,2} \cdots \vec{e}_{l,N} \rparen$ are the $l$-th eigenvalue and eigenvector, respectively. 
Note that we omitted the two translational zero modes.
The eigenvectors are orthonormalized as $\vb*{e}_l \cdot \vb*{e}_m = \delta_{lm}$, where $\delta_{lm}$ is the Kronecker delta.

For structural glasses, the eigenvalues and eigenvectors of the dynamical matrix have been extensively studied in the context of vibrational dynamics.
In this case, the dynamics are underdamped, and the square root of the eigenvalue corresponds to the frequency of vibration.
However, in our system, the equation of motion~(\ref{eqom2}) is overdamped, and hence vibrational motion is absent.
Instead, we use the eigenvalue $\lambda_l$ to determine the characteristic relaxation rate of each mode~$l$.
Nevertheless, it is still useful to introduce the square root of the eigenvalues, denoted by $\omega_l$, to help us facilitate the use of knowledge of the vibrational normal modes in structural glasses.
Therefore, we introduce
\begin{eqnarray} \label{def.frequency}
\omega_l = \sqrt{\lambda_l},
\end{eqnarray}
and call this quantity ``frequency", although it has the physical dimension of the inverse of the square root of time.
On the other hand, $\omega_l^2=\lambda_l$ has the physical dimension of the inverse of time and represents the characteristic relaxation rate of the mode~$l$.

We now expand the displacement $\vb*{u}(t)$ and the active force $\vb*{f}(t)$ using the eigenvectors as
\begin{equation} \label{eq:expand}
\begin{aligned}
\vb*{u}(t) &= \sum_{l=1}^{2N-2} c_l(t) \vb*{e}_l, \\
\vb*{f}(t) &= \sum_{l=1}^{2N-2} f_l(t) \vb*{e}_l,
\end{aligned}
\end{equation}
where $c_l(t) = \vb*{u}(t) \cdot \vb*{e}_l$ is the displacement along the mode $l$ and $f_l(t) = \vb*{f}(t) \cdot \vb*{e}_l$ is the active force acting on the mode $l$.
Substituting Eq.~(\ref{eq:expand}) into Eq.~(\ref{eqom2}), we obtain the equation of motion for each mode $l$ as 
\begin{equation} \label{eq_c_l}
\gamma \dv{c_l}{t} = - \gamma \omega_l^2 c_l + f_l \qquad (l = 1,2, \cdots, 2N-2).
\end{equation}
The steady-state solution of Eq.~(\ref{eq_c_l}) is 
\begin{equation} \label{theory_c_l}
c_l(t) = \frac{1}{\gamma} \int_{-\infty}^{t} dt' f_l(t') \exp \left[ -\omega_l^2(t-t') \right].
\end{equation}
Note again that the characteristic time scale is $\omega_l^{-2} = \lambda_l^{-1}$ rather than $\omega_l^{-1}$. 

\subsection{Observables}~\label{sec.observables}
In this work, we utilize four physical observables to characterize the fluctuation dynamics of active glasses. 
Here, we define these observables and express them in terms of the vibrational eigenmodes.

Firstly, we consider the energies associated with the eigenmodes. 
Since the stiffness of the mode $l$ is given by $\gamma \omega_l^2$, we introduce 
\begin{equation} \label{def_E}
E_l = \frac{1}{2} \gamma \omega_l^2 \langle \left( \vb*{u} \cdot \vb*{e}_l \right)^2 \rangle = \frac{1}{2} \gamma \omega_l^2 \langle c_l^2 \rangle, 
\end{equation}
which quantifies the average excitation energy of the normal mode $l$. 

Secondly, we consider the MSD as 
\begin{equation} \label{def_msd}
\begin{aligned}
R(t) &= \frac{1}{N} \langle \left| \vb*{u}(t) - \vb*{u}(0) \right|^2 \rangle, \\
 &= \frac{2}{N} \sum_{l=1}^{2N-2} \left( \langle c_l^2 \rangle - \langle c_l(t)c_l(0) \rangle \right), 
\end{aligned}
\end{equation}
which characterizes the single particle displacement in the active glasses. 

Thirdly, we focus on the spatial correlation of the displacements in the active glasses. 
Specifically, we consider the longitudinal and transverse correlation functions of the displacement field, $S_L(k)$ and $S_T(k)$, which are defined as~\cite{Ikeda_2015} 
\begin{equation} \label{def_sk}
\begin{aligned}
S_L(k) &= \frac{k^2}{N} \langle \vec{u}_{L,\vec{k}} \cdot \vec{u}_{L,-\vec{k}}\rangle, \\
S_T(k) &= \frac{k^2}{N} \langle \vec{u}_{T,\vec{k}} \cdot \vec{u}_{T,-\vec{k}}\rangle.
\end{aligned}
\end{equation}
Here, $\vec{u}_{\vec{k}}$ is the Fourier transform of the displacement field $\vb*{u} = (\vec{u}_1, \vec{u}_2, \cdots, \vec{u}_N)$, and when the displacement is sufficiently small, it is given by
\begin{equation}
\vec{u}_{\vec{k}} =  \sum_{i=1}^N \vec{u}_i \exp (-i\vec{k} \cdot  \vec{r}_{0,i} ),
\end{equation}
where $\vec{r}_{0,i}$ denotes the position of the particle $i$ in the inherent structure $\vb*{r}_{0}$.
The longitudinal and transverse parts are defined as $\vec{u}_{L,\vec{k}} = \left( \hat{\vec{k}} \cdot \vec{u}_{\vec{k}} \right)\hat{\vec{k}}$ and $\vec{u}_{T,\vec{k}} = \vec{u}_{\vec{k}} - \vec{u}_{L,\vec{k}}$, where $\hat{\vec{k}} = \vec{k}/k$ is the unit vector in the direction of $\vec{k}$, and $k = |\vec{k}|$.
Using Eq.~(\ref{eq:expand}), we can express these correlation functions in terms of the eigenmodes as
\begin{equation} \label{def_sk2}
\begin{aligned}
&S_L(k) = \frac{k^2}{N} \sum_{i,j=1}^N \sum_{l,m=1}^{2N-2} \langle c_l c_m \rangle \\
&\times \left( \hat{\vec{k}} \cdot \vec{e}_{l,i} \right) \left( \hat{\vec{k}} \cdot \vec{e}^{\dagger}_{m,j} \right) \exp\left[ i\left( \vec{k} \cdot \left( \vec{r}_{0,j} - \vec{r}_{0,i} \right) \right) \right],\\
&S_T(k) = \frac{k^2}{N} \sum_{i,j=1}^N \sum_{l,m=1}^{2N-2} \langle c_l c_m \rangle \\
&\times \left[ \vec{e}_{l,i}  - \hat{\vec{k}} \left( \hat{\vec{k}} \cdot \vec{e}_{l,i} \right) \right] \left[ \vec{e}^{\dagger}_{m,j} - \hat{\vec{k}} \left(\hat{\vec{k}} \cdot \vec{e}^{\dagger}_{m,j} \right) \right] \\
&\times \exp\left[ i\left( \vec{k} \cdot \left( \vec{r}_{0,j} - \vec{r}_{0,i} \right) \right) \right].
\end{aligned}
\end{equation}
%

It is worth noting that all of the above three observables are determined by the time correlation function $\langle c_l(t) c_m(t') \rangle$.
Using Eq.~(\ref{theory_c_l}), this time correlation is formulated as 
\begin{equation} \label{eq.corfl}
\begin{aligned}
& \langle c_l(t) c_m(t') \rangle = \frac{1}{\gamma^2} \int_{-\infty}^{t} dt_1 \int_{-\infty}^{t'} dt_2 \langle f_l(t_1) f_m(t_2) \rangle \\
& \times \exp \left[ -\omega_l^2(t-t_1) \right] \exp \left[ -\omega_m^2(t'-t_2) \right].
\end{aligned}
\end{equation}
Therefore, the time correlation of the active forces $\langle f_l(t) f_m(t') \rangle$ fully determines $\langle c_l(t) c_m(t') \rangle$, and thus the three observables introduced above.

Finally, we quantify the non-Gaussianity of the fluctuation dynamics by introducing two NGPs. 
One NGP is defined for the active force as
\begin{equation} \label{def_fNGP}
\alpha_{f_l} = \frac{\langle f_l^4 \rangle}{3 \langle f_l^2 \rangle^2}-1,
\end{equation}
which quantifies the non-Gaussianity of the active force acting on the mode $l$.
The other NGP is for the displacement as
\begin{equation} \label{def_cNGP}
\alpha_{c_l} = \frac{\langle c_l^4 \rangle}{3 \langle c_l^2 \rangle^2}-1, 
\end{equation}
which quantifies the non-Gaussianity of $c_l$, the particles' displacement along the mode $l$. 

\section{Statistical properties of active forces}~\label{sec.activenoises}
In this section, we discuss statistical properties of the active forces $f_l(t) = \vb*{f}(t) \cdot \vb*{e}_l$ in the AOUP, ABP, and RTP glasses. 
Specifically, for each active glass, we calculate the time correlation $\langle f_l(t) f_m(t') \rangle$ and discuss the non-Gaussianity of the probability distribution of $f_l$, $P(f_l)$. 

\subsection{AOUP glass}~\label{sec.activenoises.aoup}
Projecting Eq.~(\ref{eq:aoup}) onto $\vb*{e}_l$, we can obtain the equation of motion for $f_l(t)$ as
\begin{equation}
 \label{aoupforce}
\tau \dv{f_l}{t} = - f_l + \sqrt{2 \gamma D} \eta_l,
\end{equation}
where $\eta_l =  \vb*{\eta} \cdot \vb*{e}_l$ is the Gaussian white noise which satisfies $\langle \eta_l(t) \rangle = 0$ and $\langle \eta_l(t) \eta_l(t') \rangle =\delta(t-t')$.
As a result, we can determine the time correlation as
\begin{equation}
\begin{aligned} \label{noise_correlation_2}
\langle f_l(t) f_m(t') \rangle = \frac{\gamma D}{\tau}\delta_{lm} \exp \left( -\frac{|t-t'|}{\tau} \right).
\end{aligned}
\end{equation}
Moreover, Eq.~(\ref{aoupforce}) indicates that the probability distribution $P(f_l)$ is Gaussian.
Therefore, the corresponding NGP is zero, $\alpha_{f_l}=0$.

\subsection{ABP glass}
For the ABP glass, we first project Eq.~(\ref{eq:abp}) onto $\vb*{e}_l$ to obtain 
\begin{equation} \label{f_l_ABP}
f_l(t) = \gamma v_0 \sum_{i=1}^{N} \left[ e_{l,i}^x \cos{\theta_i}(t) + e_{l,i}^y \sin{\theta_i}(t) \right],
\end{equation}
where we introduce the notation $\vec{e}_{l,i} = ( e^x_{l,i}, e^y_{l,i} )$ for the eigenvectors. 
Using Eq.~(\ref{f_l_ABP}), we can expand $\langle f_l(t) f_m(t') \rangle$ as 
\begin{equation} \label{noise_corr}
\begin{aligned}
\langle f_l(t) f_m(t') \rangle =& \gamma^2 v_0^2 \sum_{i=1}^{N} \sum_{j=1}^{N} \lbrace e_{l,i}^x e_{m,j}^x \langle \cos{\theta_i(t)} \cos{\theta_j(t')} \rangle \\ 
&+e_{l,i}^x e_{m,j}^y \langle \cos{\theta_i(t)} \sin{\theta_j(t')} \rangle \\ 
&+e_{l,i}^y e_{m,j}^x \langle \sin{\theta_i(t)} \cos{\theta_j(t')} \rangle \\ 
&+e_{l,i}^y e_{m,j}^y \langle \sin{\theta_i(t)} \sin{\theta_j(t')} \rangle \rbrace.\\
\end{aligned}
\end{equation}
The time correlations on the right-hand side of Eq.~(\ref{noise_corr}) can be evaluated as follows. 
Since $\vb*{\theta}(t)$ follows the Wiener process Eq.(\ref{eq.abp_theta}), the conditional probability $P(\vb*{\theta}, t | \vb*{\theta}', t')$, which expresses the probability that the angle vector is $\vb*{\theta}$ at time $t$ given that it is $\vb*{\theta}'$ at time $t'$, is expressed as~\cite{Gardiner2009} 
\begin{equation}
P(\vb*{\theta}, t | \vb*{\theta}', t') = \frac{1}{ \left( 4 \pi D_R\left| t-t' \right| \right)^{N/2}} \prod_{i=1}^{N} \exp \left\{ -\frac{(\theta_i - \theta'_i)^2}{4D_R \left| t-t' \right|} \right\}. 
\end{equation}
Using this conditional probability, the time correlations can be calculated by conducting the Gaussian integrals, \textit{e.g.}, 
\begin{equation} \label{expected value}
\begin{aligned}
&\langle \cos{\theta_i(t)} \cos{\theta_j(t')} \rangle \\
& = \int d\vb*{\theta}  \int d\vb*{\theta}' P(\vb*{\theta}, t | \vb*{\theta}', t') P(\vb*{\theta}', t') \cos{\theta_i} \cos{\theta'_j} , \\
& = \frac{1}{2} e^{-D_R| t-t' |} \delta_{ij}.
\end{aligned}
\end{equation}
In the second line in Eq.~(\ref{expected value}), we introduce $P(\vb*{\theta'}, t')$ to be the probability distribution of $\vb*{\theta}'$ at time $t'$, and in the final line, we assume $P(\vb*{\theta'}, t')$ to be a uniform distribution as we focus on the steady state. 
Similar calculations lead to
\begin{equation}
\begin{aligned}
\langle \sin{\theta_i(t)} \sin{\theta_j(t')} \rangle &= \frac{1}{2} e^{-D_R| t-t' |} \delta_{ij}, \\
\langle \cos{\theta_i(t)} \sin{\theta_j(t')} \rangle &= 0.
\end{aligned}
\end{equation}
Inserting them into Eq.~(\ref{noise_corr}), we arrive at 
\begin{equation} \label{noise_correlation_2a}
\begin{aligned}
&\langle f_l(t) f_m(t') \rangle \\
& = \gamma^2 v_0^2 \sum_{i=1}^{N} \sum_{j=1}^{N} \lbrace e_{l,i}^x e_{m,j}^x + e_{l,i}^y e_{m,j}^y \rbrace   \frac{1}{2}  e^{-D_R|t-t'|} \delta_{ij}, \\
&= \frac{\gamma^2 v_0^2}{2}\delta_{lm} \exp \left( -D_R|t-t'| \right),
\end{aligned}
\end{equation}
where we use the orthonormal condition of the eigenmodes in calculating from the second line to the final line.
Therefore, the time correlations of the active forces in the ABPs decay exponentially, as in the case of the AOUPs. 
This is valid for any eigenmode, regardless of extended or localized mode. 

We next discuss the non-Gaussianity of the probability distribution $P(f_l)$. 
Although the active force acting on a single particle, $\vec{f}_i$, is highly non-Gaussian in the ABPs, we will conclude that $f_l$ becomes a Gaussian random variable if the eigenvector $\vb*{e}_l$ is spatially extended.
We start with the following expression for the eigenmode;
\begin{equation} \label{spatial_extended}
\vb*{e}_l = \frac{1}{\sqrt{N}} ( a_{l,1}, b_{l,1}, \cdots, a_{l,N}, b_{l,N} ).  
\end{equation}
If this mode is spatially extended, each particle carries a comparable contribution; thus, $a_{l,i}^2 + b_{l,i}^2 \approx 1$ is valid for any particle $i$. 
Then, $f_l$ is evaluated as
\begin{equation} \label{f_l_extended}
f_l = \vb*{f} \cdot \vb*{e}_{l} \approx \frac{\gamma v_0}{\sqrt{N}} \sum_{i=1}^N \cos\left( \theta_i - \beta_{l,i} \right),
\end{equation}
where $\tan\beta_{l,i} =b_{l,i}/ a_{l,i}$. 
As $\theta_i$ is a random variable, $\cos{(\theta_i - \beta_{l,i})}$ is also a random variable.
Although this variable is not Gaussian, the central limit theorem guarantees that the right-hand side of Eq.~(\ref{f_l_extended}) will be a Gaussian random variable when $N \to \infty$. 
This indicates that for any spatially extended mode $l$, the distribution of the active force $P(f_l)$ will be Gaussian, and thus $\alpha_{f_l} = 0$.
This conclusion is valid for any spatially extended mode, including both the phonon-like mode and the disordered extended mode, as the argument requires only the random nature of $\theta_i$~(not $\beta_{l,i}$). 
However, this argument cannot be applied to the spatially localized eigenmodes, and their distribution of the active force $P(f_l)$ may not be Gaussian.

\subsection{RTP glass}
The active force $f_l$ in the RTP glass shares the same expression Eq.~(\ref{f_l_ABP}) with the ABP glass. 
Therefore, the time correlations can be written as in Eq.~(\ref{noise_corr}). 
A difference between the RTPs and ABPs comes from the time evolution of $\vb*{\theta}(t)$. 
In the RTP glass, $\vb*{\theta}(t)$ follows the Poisson process, and the conditional probability is given by~\cite{Gardiner2009,Santra_2020}
\begin{equation}
\begin{aligned}
&P(\vb*{\theta},t | \vb*{\theta}',t') = \\
& \prod_{i=1}^{N} \left\{ e^{-R_{\rm tumble} |t-t'|} \delta(\theta_i - \theta'_i) + \left[ 1 - e^{-R_{\rm tumble} |t-t'|} \right] \frac{1}{2 \pi} \right\}.
\end{aligned}
\end{equation}
Plugging this into the second line of Eq.~(\ref{expected value}) and assuming the steady state condition, we obtain
\begin{equation}
\begin{aligned}
\langle \cos{\theta_i(t)} \cos{\theta_j(t')} \rangle &= \frac{1}{2}e^{-R_{\rm tumble} | t-t' |} \delta_{ij},\\
\langle \sin{\theta_i(t)} \sin{\theta_j(t')} \rangle &= \frac{1}{2}e^{-R_{\rm tumble} | t-t' |} \delta_{ij},\\
\langle \cos{\theta_i(t)} \sin{\theta_j(t')} \rangle &= 0.
\end{aligned}
\end{equation}
Inserting these results into Eq.~(\ref{noise_corr}), we obtain 
\begin{equation} \label{noise_correlation_2b}
\langle f_l(t) f_m(t') \rangle = \frac{\gamma^2 v_0^2}{2}\delta_{lm} \exp \left( -R_{\rm tumble} |t-t'| \right).
\end{equation}
As in the case of the ABP glass, this result is valid for all types of eigenmodes. 

For the non-Gaussianity, we can make the same argument as in the ABP glass.
This suggests that for a mode $l$ that is spatially extended, the probability distribution of $f_l$ is Gaussian, and its NGP, $\alpha_{f_l}$, equals zero.
However, like in the ABP glass, this argument does not restrict non-Gaussianity for modes that are spatially localized.

\subsection{Comparison of three active forces}
%
\begin{table}[t]
\tabcolsep = 0.4cm
\renewcommand{\arraystretch}{2.25}
\caption{Typical amplitude $\langle f_l^2 \rangle$ and the persistence time of the active forces in three different active glasses.}
\label{tab_T,tau}
\centering
\begin{tabular}{c||c|c|c}
\quad & AOUPs & ABPs & RTPs  \\
\hline
\hline
\mbox{Amplitude} & $\gamma D/\tau$ & $\gamma^2 v_0^2/2$ & $\gamma^2 v_0^2/2$ \\
\hline
\mbox{Persistence time} & $\tau$ & ${1}/{D_R}$ & ${1}/{R_{\rm tumble}}$ \\
\hline
\end{tabular}
\end{table}
%
So far, we obtained the time correlations of three active forces; Eq.~(\ref{noise_correlation_2}) for the AOUPs, Eq.~(\ref{noise_correlation_2a}) for the ABPs, and Eq.~(\ref{noise_correlation_2b}) for the RTPs, all of them are similar expressions.
We summarize the typical amplitude $\langle f_l^2 \rangle$ and the persistence time of the three active forces in Table~\ref{tab_T,tau}.
If the amplitude and persistence time of the three active forces are the same, their time correlations become identical.
It is noteworthy that $f_l$ behaves as a Gaussian random variable not only in the AOUP glass but also in the ABP and RTP glasses for the spatially extended mode $l$.

Moving forward, we will utilize the parameters $(D, \tau)$ to identify the three active forces.
Specifically, $(v_0=\sqrt{2D/\tau \gamma}, D_R=1/\tau)$ is used to describe the ABP glass, while $(v_0=\sqrt{2D/\tau \gamma}, R_{\rm tumble}=1/\tau)$ is used to describe the RTP glass. This parameterization will enable us to compare the three active glasses on an equal footing of $(D,\tau)$.

\section{Numerical model}~\label{sec.Numericalmodel}
%
\begin{figure}[t]
\centering
\includegraphics[width=80mm]{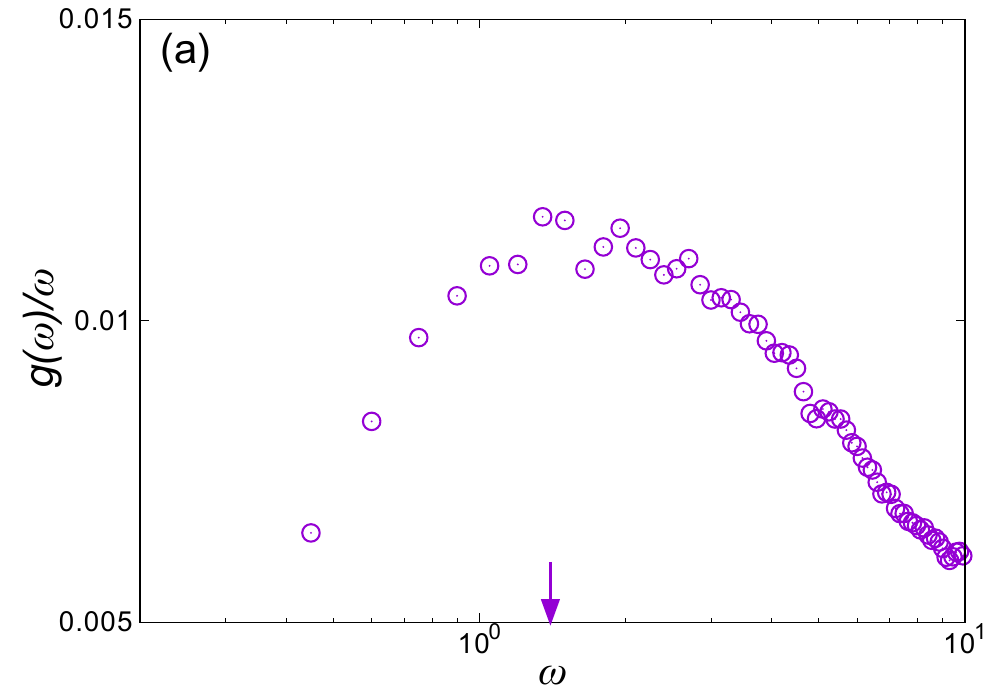}
\includegraphics[width=80mm]{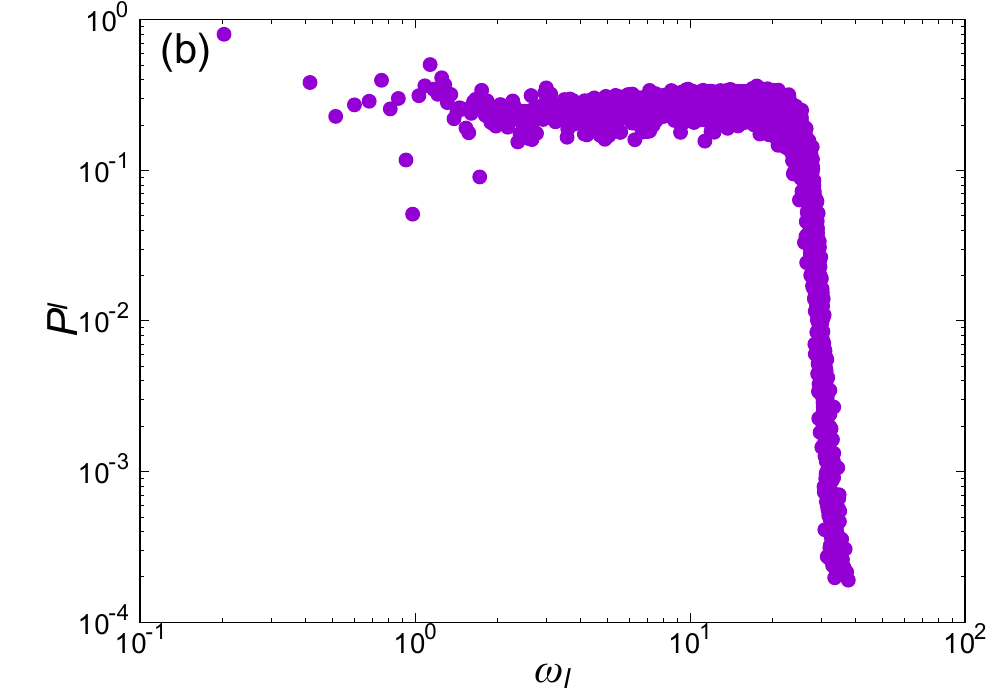}
\caption{
(a) The reduced vDOS $g(\omega)/ \omega$ against frequency $\omega$.
(b) The participation ratio $P^l$ against frequency $\omega_l$ for each mode $l$.
The vertical arrow in (a) indicates the frequency of the boson peak $\omega_\text{BP}$.
}
\label{fig.vib}
\end{figure}
%
The formulations we have developed in the previous section enable us to calculate the observables when the dynamical matrix $\mathcal{M}$ is specified. 
Here we introduce our numerical model for $\mathcal{M}$. 

We consider a 50:50 binary mixture of large and small particles in a $2$-dimensional box with the dimension $L$. 
The diameters of small and large particles are denoted by $\sigma$ and $\sigma'$, respectively, with the size ratio $\sigma'/\sigma = 1.4$. 
The total potential is given by ${U}(\vb*{r}) = \sum_{i>j} v_{ij}(r_{ij})$, where $v_{ij}(r_{ij})$ is the WCA interparticle interaction,
\begin{equation}
v_{ij}(r_{ij}) = \epsilon \left[ \left( \frac{r_{ij}}{\sigma_{ij}} \right) ^{-12} - \left( \frac{r_{ij}}{\sigma_{ij}} \right) ^{-6} + \frac{1}{4} \right] \Theta(r_{ij} - r_{ij}^c). 
\end{equation}
$r_{ij} = |\vec{r}_j - \vec{r}_i|$ is the distance between the particles $i$ and $j$, $\sigma_{ij} = {(\sigma_i + \sigma_j)}/{2}$ is the mean diameter of two particles with $\sigma_i$ being the diameter of the particle $i$.
The Heaviside step function, $\Theta(r_{ij} - r_{ij}^c)$, implements a cutoff of the potential at $r_{ij}^c = 2^{\frac{1}{6}}\sigma_{ij}$, so that the interaction is purely repulsive.  
We define the packing fraction of the system as $\varphi = \sum_{i=1}^N \pi \left({\sigma_{i}^\text{eff}}/{2} \right)^2/L^2$, where $\sigma_{i}^\text{eff} = 2^{\frac{1}{6}}\sigma_{i}$ represents an ``effective" diameter of the particle $i$. 
The effective diamter satisfies $r_{ij}^c = (\sigma_{i}^\text{eff} + \sigma_{j}^\text{eff})/2$ to ensure that the particle volume $\pi \left({\sigma_{i}^\text{eff}}/{2}\right)^2$ is compartible with the actual excluded volume. 
We set $N=16000$ and tune $L$ such that $\varphi= 1.007$, which is well above the jamming density. 
In the following, we use $\sigma$, $\epsilon$, and $\gamma \sigma^2/ \epsilon$ as the units of length, energy, and time, respectively.

We begin by creating a random configuration in the simulation box.
We next use the Fast Inertial Relaxation Engine (FIRE) algorithm~\cite{Bitzek_2006} to obtain a mechanically stable configuration represented by $\vb*{r}_{0}$ and the associated dynamical matrix $\mathcal{M}$ defined in Eq.~(\ref{dmat}).
We then perform numerical diagonalization of the matrix $\mathcal{M}$ to determine the eigenfrequencies and eigenvectors~\cite{MizunoIkeda2022}.
Employing these quantities, we calculate the physical observables using formulas obtained in previous sections.
For the parameters of the active forces, we set $D=0.1$ and vary the persistence time $\tau$.
It is important to note that the selection of $D$ is arbitrary since the equation of motion, Eq.~(\ref{eqom2}), is linear.
We chose $D=0.1$ based on our observation that the plateau height of the MSD computed below is comparable to experimental data of bacterial glasses~\cite{Lama_2023}.

In addition, we conduct numerical simulations by solving the linearized equation of motion, Eq.~(\ref{eqom2}), with three different active forces, all of which begin from the same mechanically stable configuration $\vb*{r}_{0}$.
We adopted the Predictor-Corrector Method for the numerical integration.
The active forces were computed using different methods depending on the type of system.
For the AOUP glass, we use the Gillespie method~\cite{Gillespie_1996}, while for the ABP glass, we use the Euler-Maruyama method.
For the RTP glass, we select the update event of the active forces to occur with a probability $P_\text{tumble} = 1 - e^{-R_\text{tumble}\Delta t}$ at each time step $\Delta t$, and then the direction of the active forces was chosen according to the uniform distribution.
In all three cases, we perform simulations for a sufficiently long period to achieve stationary states before running the production runs.

Before analyzing the structural fluctuations, we briefly discuss the nature of normal modes in the present system~(WCA glass). 
We define the density of states as~\cite{MizunoIkeda2022}
\begin{equation}
g(\omega) = \frac{1}{2N} \sum_l \delta (\omega - \omega_l).
\end{equation}
%
Figure~\ref{fig.vib}(a) shows the reduced density of states $g(\omega)/\omega$.
The boson peak is located at the frequency $\omega_\text{BP} \approx 1.4$.
It is well understood that the normal modes at $\omega \ll \omega_\text{BP}$ are plane-wave-like, while those at $\omega \gg \omega_\text{BP}$ are disordered and extended~\cite{Mizuno_2017}.

In addition, figure~\ref{fig.vib}(b) shows the participation ratio $P^l$ of each normal mode $l$, which is defined as~\cite{MizunoIkeda2022}
\begin{equation}
P^l = \frac{1}{N} \left[ \sum_{i=1}^N \left(\vec{e}_{l,i} \cdot \vec{e}_{l,i} \right)^2 \right]^{-1}.
\end{equation}
$P^l$ measures the fraction of particles that participate in the vibrations.
As extreme cases, $P^l = 1$ for an ideal mode in which all constituent particles vibrate equally, while $P^l = 1/N \ll 1$ for an ideal mode involving only one particle.
The normal modes at the high-frequency edge are spatially localized, while most of the other modes are spatially extended.
Notably, in the low-frequency regime, there are only a few quasi-localized modes, which is consistent with previous studies on the vibrations of two-dimensional packings~\cite{Mizuno_2017}.

\section{Fluctuation dynamics in the active glasses}
We will now analyze the fluctuation dynamics in active glasses using the physical observables that were introduced and formulated above.

\subsection{Time correlation function of $c_l$}~\label{sec.timecorrelationcl}
In Sec.~\ref{sec.observables}, we demonstrated that the time correlation function of the displacements along the modes, denoted by $\langle c_l(t) c_m(t') \rangle$, plays a central role in determining the observables.
The closed form of $\langle f_l(t) f_m(t') \rangle$ that we obtained in Sec.~\ref{sec.activenoises} allows us to calculate $\langle c_l(t) c_m(t') \rangle$.
Specifically, by substituting Eq.~(\ref{noise_correlation_2}) into Eq.~(\ref{eq.corfl}), we obtain the following expression;
\begin{equation}
\begin{aligned} \label{eq.corfl2}
& \langle c_l(t) c_m(t') \rangle = \delta_{lm} \frac{D}{\gamma} \bigg\{ \frac{\tau}{\tau \omega_l^2 \left( \tau \omega_l^2 + 1 \right) } e^{-\omega_l^2 |t-t'|}  \\
&+ \frac{\tau}{ \left(\tau \omega_l^2 - 1 \right) \left(\tau \omega_l^2 + 1 \right) } \left[ e^{- |t-t'|/\tau} - e^{-\omega_l^2 |t-t'|} \right] \bigg\}.
\end{aligned}
\end{equation}
This result is applicable to all of the AOUP, ABP, and RTP glasses, as $\langle f_l(t) f_m(t') \rangle$ is the same among these models after the reparametrization of the active forces (refer to Table~\ref{tab_T,tau}). 

It can be insightful to examine Eq.~(\ref{eq.corfl2}) in two extreme scenarios.
The first one is when the persistence time is far shorter than the characteristic relaxation time of the mode $l$, \textit{i.e.}, $\tau \ll \omega_l^{-2}$.
In this case, Eq.~(\ref{eq.corfl2}) can be simplified to
\begin{equation} \label{eq.corfl3}
\langle c_l(t) c_l(t') \rangle \simeq \frac{D}{\gamma \omega_l^2} \exp\left( -\omega_l^2 |t-t'| \right).
\end{equation}
This outcome is identical to the one for thermal glasses~\cite{Coslovich2022}, where $D$ denotes temperature and $\gamma \omega_l^2$ symbolizes the stiffness of the mode $l$.
It demonstrates that under the condition $\tau \ll \omega_l^{-2}$ which we term the ``thermal condition" from now on, the active forces function like a random white noise for the mode $l$, and the dynamics along it are equivalent to those in thermal systems.

The other extreme case is that the persistence time is much longer than the relaxation time, \textit{i.e.}, $\tau \gg \omega_l^{-2}$, and Eq.~(\ref{eq.corfl2}) can be simplified to 
\begin{equation} \label{eq.corfl4}
\langle c_l(t) c_l(t') \rangle \simeq \frac{D}{\gamma \tau \omega_l^4} \exp\left( - \frac{|t-t'|}{\tau} \right).
\end{equation}
In this scenario, the time correlation of $c_l$ decays with the time scale $\tau$ of $f_l$.
This suggests that the active forces are effectively equivalent to a constant force.
Therefore, the equation of motion of $c_l$, as expressed in Eq.~(\ref{eq_c_l}), can be seen as a force-balance equation between the active forces and the restoring forces,
\begin{equation} \label{eq.fbalance}
f_l \sim \gamma \omega_l^2 c_l. 
\end{equation}
This relation with Eq.~(\ref{noise_correlation_2}) reproduces Eq.~(\ref{eq.corfl4}).  
This means that under the condition $\tau \gg \omega_l^{-2}$ which we term the ``quasi-static condition", the dynamics along the mode $l$ is quasi-static, and $c_l$ simply follows $f_l$. 

It is important to note that the typical amplitude of $c_l$ is heavily influenced by these conditions, as discussed in detail later.  
In the thermal condition, we observe that $\langle c_l^2 \rangle \propto \omega_l^{-2}$, which is consistent with the equipartition law.
However, in the quasi-static condition, $\langle c_l^2 \rangle \propto \omega_l^{-4}$, meaning that displacements along the high-frequency modes are significantly suppressed.

\subsection{Average excitation energy}
%
\begin{figure}[t]
\centering
\includegraphics[width=80mm]{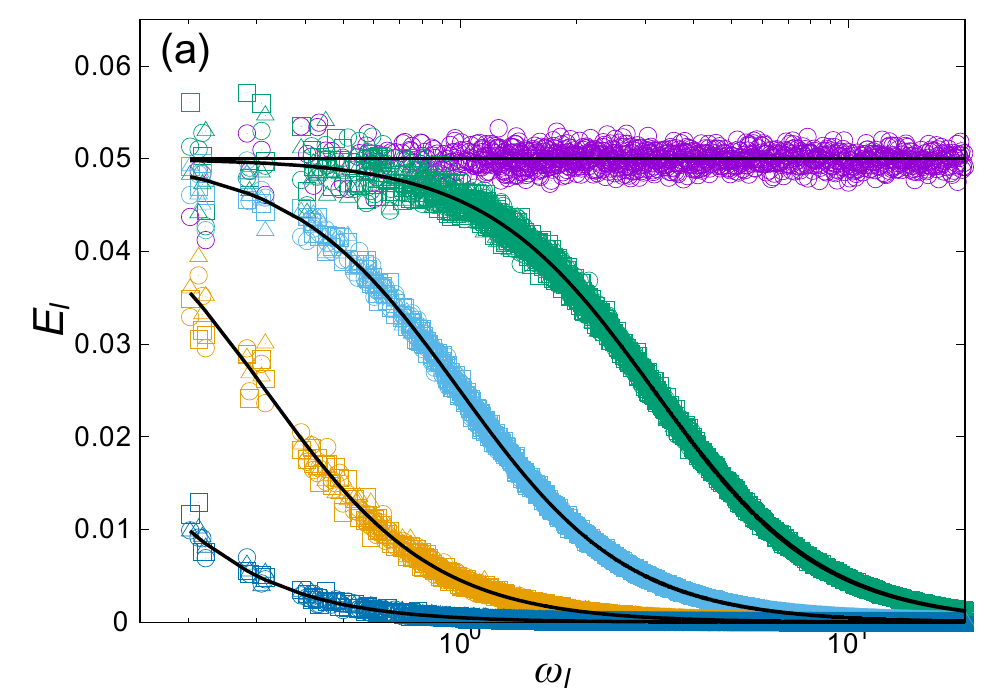}
\includegraphics[width=80mm]{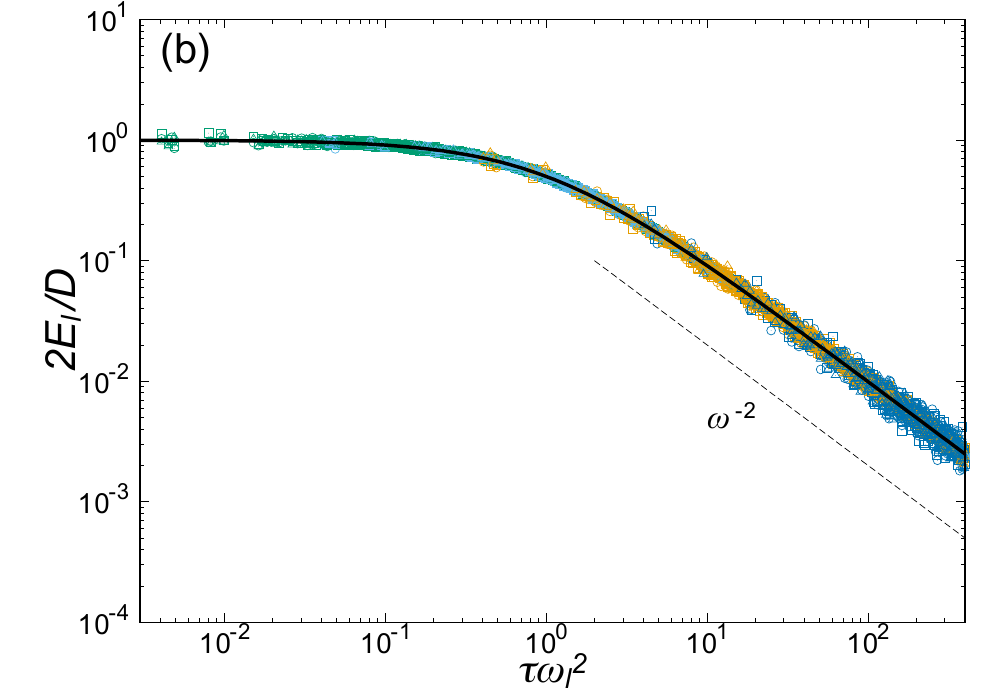}
\caption{
The average excitation energy of three active glasses.
(a) $E_l$ is plotted as a function of eigenfrequency $\omega_l$ for various persistent times $\tau$; $\tau = 10^{-8}$~(purple), $10^{-1}$~(green), $10^0$~(cyan), $10^1$~(yellow), and $10^2$~(blue).
(b) $2 E_l / D$ is plotted as a function of $\tau \omega^2_l$.
The symbols represent simulation data; circles for AOUP glass, squares for ABP glass, and triangles for RTP glass.
The solid lines indicate the theoretical prediction of Eq.~(\ref{theory_E}), which is commonly shared by the three active glasses.
The dashed line in (b) indicates $\propto \omega^{-2}$.
}
\label{fig.Ene}
\end{figure}

We now discuss the average excitation energy $E_l$.
In our numerical simulations, we calculated the time series of $c_l = \vb*{u} \cdot \vb*{e}_l$ for each mode $l$ and evaluated $E_l$ using the formula given in Eq.~(\ref{def_E}).
To illustrate our findings, we plot the results for five different persistent times $\tau=10^{-8}$, $10^{-1}$, $10^{0}$, $10^{1}$, and $10^{2}$ in Fig.~\ref{fig.Ene}(a).
The circles, squares, and triangles in the plot represent the results for the AOUP, ABP, and RTP glasses, respectively.
Notably, these results are quantitatively identical to each other for all values of $\tau$. 
When the persistence time is extremely short~($\tau = 10^{-8}$), $E_l$ remains almost constant for all modes, which is consistent with the equipartition law in thermal systems. However, as $\tau$ increases, the excitation energies of the high-frequency modes progressively decrease.

We can now rationalize numerical observations.
By substituting the expression in Eq.~(\ref{eq.corfl2}) into Eq.~(\ref{def_E}), we can derive the theoretical formula of the average excitation energy as
\begin{equation} \label{theory_E}
E_l = \frac{D}{2} \left( \frac{1}{\tau \omega_l^2 + 1} \right),
\end{equation}
which is plotted as solid lines in Fig.~\ref{fig.Ene}(a) and accurately explains the simulation results.
This formula is applicable to all three active glasses and thus explains the agreement among the outcomes of these glasses.

We discuss the theoretical formula, Eq.~(\ref{theory_E}), in the low- and high-frequency regimes, separately.  
In the low-frequency regime where $\omega_l^2$ is much smaller than $\tau^{-1}$, which is the thermal condition as discussed in Sec.~\ref{sec.timecorrelationcl}, Eq.~(\ref{theory_E}) can be simplified to
\begin{equation} \label{theory_E2}
E_l \simeq \frac{D}{2}.
\end{equation}
This equation represents the equipartition law, which is reasonable because the active force behaves like thermal noise in this frequency regime.
On the other hand, in the high-frequency regime where $\omega_l^2$ is much greater than $\tau^{-1}$, which is the quasi-static condition as discussed in Sec.~\ref{sec.timecorrelationcl}, Eq.~(\ref{theory_E}) can be simplified to
\begin{equation} \label{theory_E3}
E_l \simeq \frac{D}{2 \tau \omega_l^2}.
\end{equation}
This equation reflects the quasi-static nature of the dynamics in this frequency regime.
The active force balances with the restoring force, which is described as $f_l \sim \gamma \omega_l^2 c_l$~(Eq.~(\ref{eq.fbalance})).
As a result, the excitation energy can be estimated as
\begin{equation}
E_l = \frac{1}{2}\gamma \omega_l^2 \langle c_l^2 \rangle \sim \frac{\langle f_l^2 \rangle}{2 \gamma \omega_l^{2}} = \frac{D}{2 \tau \omega_l^{2}},
\end{equation}
which reproduces Eq.~(\ref{theory_E3}).

To precisely test the theoretical formula Eqs.~(\ref{theory_E}), (\ref{theory_E2}) and (\ref{theory_E3}), we present a scaled data plot of $2E_l/D$ against $\tau \omega_l^2$ in Fig.~\ref{fig.Ene}(b).
All the data of different persistent times and active glasses collapse onto a single curve, fully agreeing with the theory.

\subsection{MSD}
%
\begin{figure}[t]
\centering
\includegraphics[width=80mm]{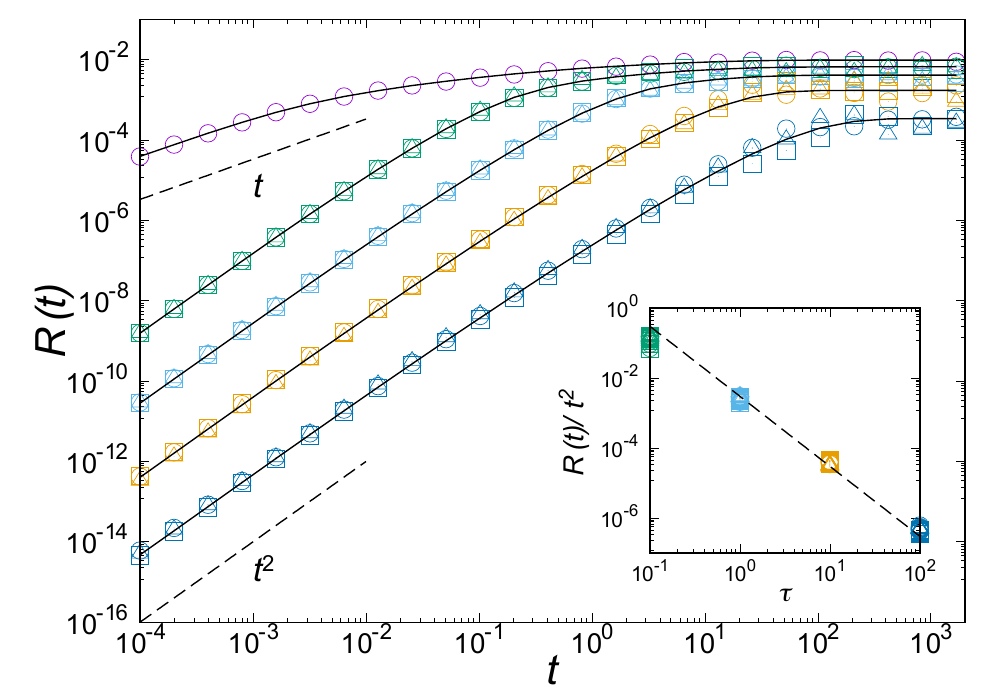}
\caption{
The MSD in three active glasses.
$R(t)$ is plotted as a function of time $t$ for various persistent times $\tau$; $\tau = 10^{-8}$~(purple), $10^{-1}$~(green), $10^0$~(cyan), $10^1$~(yellow), and $10^2$~(blue).
The symbols represent simulation data; circles for AOUP glass, squares for ABP glass, and triangles for RTP glass.
The solid lines indicate the theoretical prediction of Eq.~(\ref{theory_MSD}).
The dashed lines indicate $\propto t^2$~(ballistic behavior) and $\propto t$~(diffusive behavior).
The inset shows $R(t)/t^2$ in the ballistic regime plotted against $\tau$, indicating $R(t)/t^2 \propto \tau^{-2}$ for three active glasses.
}
\label{fig.MSD}
\end{figure}

We next discuss the MSD, denoted by $R(t)$.
To calculate $R(t)$ through numerical simulations, the standard method was used via the first line of Eq.~(\ref{def_msd}).
The results for the three active glasses with various persistent times are depicted in Fig.~\ref{fig.MSD}. 
Just like the excitation energy, the results of all three active glasses are identical.
When $\tau = 10^{-8}$, the MSD is diffusive as $R(t) \propto t$ in a short time, and it converges to a plateau in a long time, which is consistent with the known results for the thermal glasses. 
However, at finite persistence time $\tau$, the MSD becomes ballistic as $R(t) \propto t^2$ in a short time.
This observation is reasonable as the active particles try to run at a constant velocity.
The inset to Fig.~\ref{fig.MSD} demonstrates that as $\tau$ increases, the coefficient to $t^2$ in the ballistic regime, \textit{i.e.}, $R(t)/t^2$, decreases as $\tau^{-2}$ in the large $\tau$ region.
Concomitantly, the long-time plateau becomes lower as $\tau$ increases.

We can use Eq.~(\ref{eq.corfl2}) in Eq.~(\ref{def_msd}) to obtain the theoretical expression for MSD as follows;
\begin{equation} \label{theory_MSD}
\begin{aligned}
& R(t) = \frac{2D}{\gamma N} \sum_{l=1}^{2N-2} \bigg\{ \frac{\tau}{\tau \omega_l^2 \left( \tau \omega_l^2 + 1 \right) } \left( 1- e^{-\omega_l^2 t} \right) \\
&- \frac{\tau}{ \left(\tau \omega_l^2 - 1 \right) \left(\tau \omega_l^2 + 1 \right) } \left( e^{- t/\tau} - e^{-\omega_l^2 t} \right) \bigg\}.
\end{aligned}
\end{equation}
In Fig.~\ref{fig.MSD}, we plotted this formula as solid lines, which agree well with the simulation results.
It is also important to note that this formula is valid for all three active glasses.

To discuss the short-time dynamics, we expand Eq.~(\ref{theory_MSD}) with respect to $t$. 
For finite $\tau$, the leading contribution is the ballistic behavior, given by
\begin{equation} \label{theory_MSD2}
R(t) \simeq A v_0^2 t^2  \quad \text{with} \quad A=\frac{1}{2N} \sum_{l=1}^{2N-2} \frac{1}{\tau \omega_l^2 + 1},
\end{equation}
where $v_0 = \sqrt{2D/(\gamma \tau)}$ is the self-propelled velocity of non-interacting active particles (see Table.~\ref{tab_T,tau}). 
Eq.~(\ref{theory_MSD2}) indicates that the velocity of the particle in the active glasses is effectively $\sqrt{A} v_0$, which differs from the velocity of the non-interacting active particles, \textit{i.e.}, $v_0$.
This is in contrast to thermal glasses, where the short-time dynamics are always consistent with those of the non-interacting particles.
As $A$ is always less than $1$ for any $\tau$, the motions of particles become slower in the active glasses.
This can be understood based on our previous discussions. 
For the high-frequency modes $\omega_l^2 \gg \tau$, the dynamics is quasi-static. 
This means that the force balance between the active force and the restoring force is achieved so that the motion along the mode $l$ virtually stops and cannot contribute to the MSD.
The coefficient $A$ is considered as the correction arising from such modes.

We can now explain the observation in Fig.~\ref{fig.MSD} that $R(t)/t^2 \propto \tau^{-2}$ in the large $\tau$ region by using $R(t)/t^2 = A v_0^2$ in Eq.~(\ref{theory_MSD2}).
There are two effects to consider.
The first is that $v_0^2$ decays as $\tau^{-1}$ when $D$ is fixed.
The second effect is non-trivial, as $A$ decays as $\tau^{-1}$, which comes from the fact that the longer the persistence time is, the more quasi-static the modes become.
Even if we choose $(v_0,\tau)$ as the control parameters of the active forces, the second effect still remains, even though the first effect disappears.

We finally discuss the plateau height of the MSD. 
By taking the long time limit of Eq.~(\ref{theory_MSD}), we can derive
\begin{equation} \label{theory_MSD3}
R(t \to \infty) \simeq \frac{2D}{\gamma N} \sum_{l=1}^{2N-2} \frac{1}{\omega_l^2 \left( \tau \omega_l^2 + 1 \right)}. 
\end{equation}
If we set $\tau = 0$, the well-known result for the thermal glasses is reproduced, \textit{i.e.}, $R(t \to \infty) \simeq \frac{2D}{\gamma N} \sum_{l=1}^{2N-2} \frac{1}{\omega_l^2}$. 
The additional factor $1/( \tau \omega_l^2 + 1 )$ expresses the impact of activeness, which again comes from the quasi-static nature of the high-frequency modes. 
Since the displacements along such modes are strongly suppressed, the plateau height becomes lower. 

\subsection{Spatial correlation of displacements}
%
\begin{figure}[t]
\centering
\includegraphics[width=80mm]{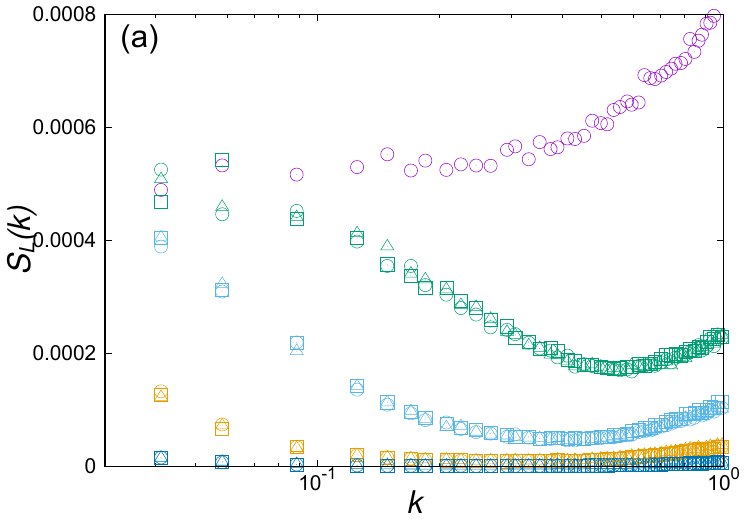}
\includegraphics[width=80mm]{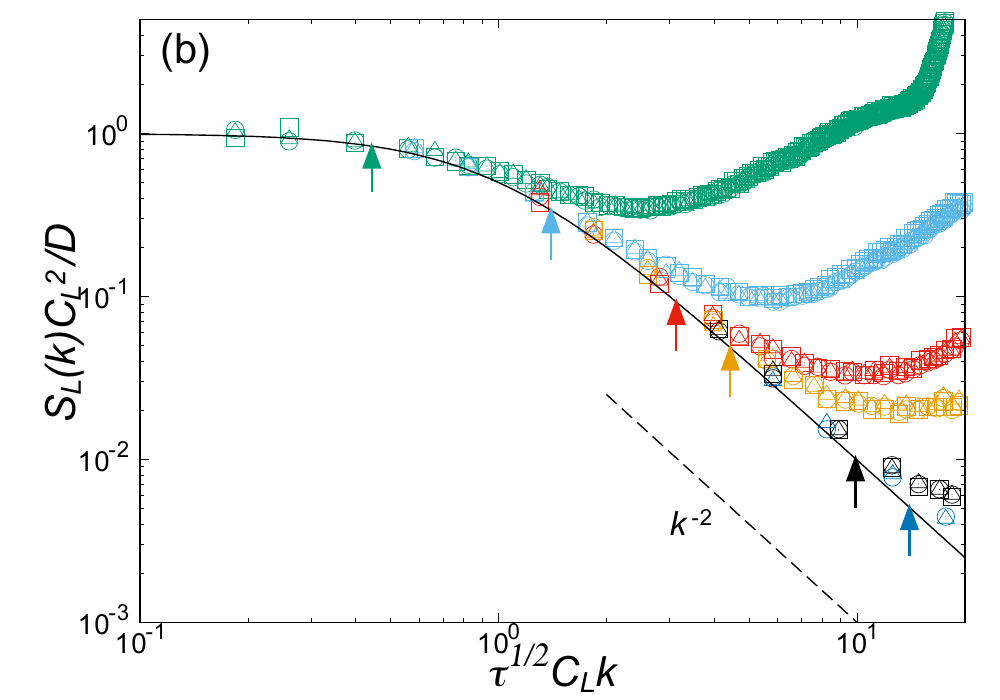}
\caption{
Correlation function of the longitudinal displacements.
(a) $S_L(k)$ is plotted as a function of wavenumber $k$ for various persistent times $\tau$; $\tau = 10^{-8}$~(purple), $10^{-1}$~(green), $10^0$~(cyan), $5 \times 10^0$~(red), $10^1$~(yellow), $5 \times 10^1$~(black), and $10^2$~(blue).
(b) $S_L(k) C_L^2 / D$ is plotted as a function of $\sqrt{\tau} C_L k$. 
The symbols represent simulation data; circles for AOUP glass, squares for ABP glass, and triangles for RTP glass.
The solid lines indicate the theoretical prediction in Eq.~(\ref{theory_Sk_scale}) with $\alpha =L$.
In (b), the vertical arrows indicate $\tau^{1/2} \omega_\text{BP}$, and the dashed line indicates $\propto k^{-2}$.
}
\label{fig.SLk}
\end{figure}

\begin{figure}[t]
\centering
\includegraphics[width=80mm]{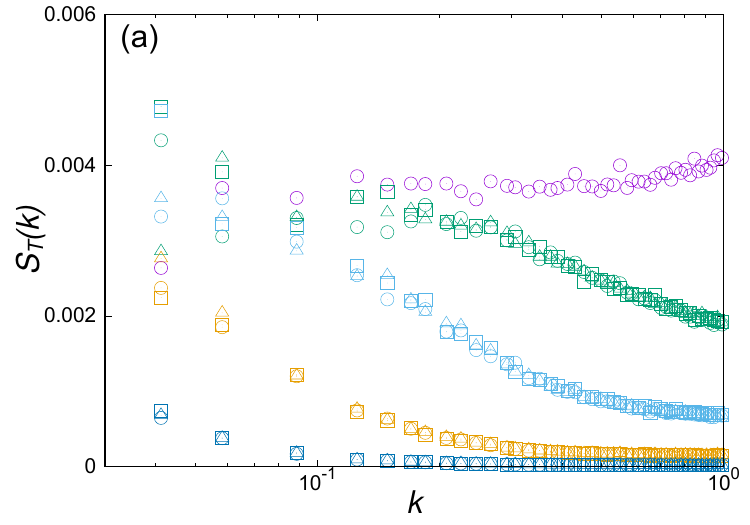}
\includegraphics[width=80mm]{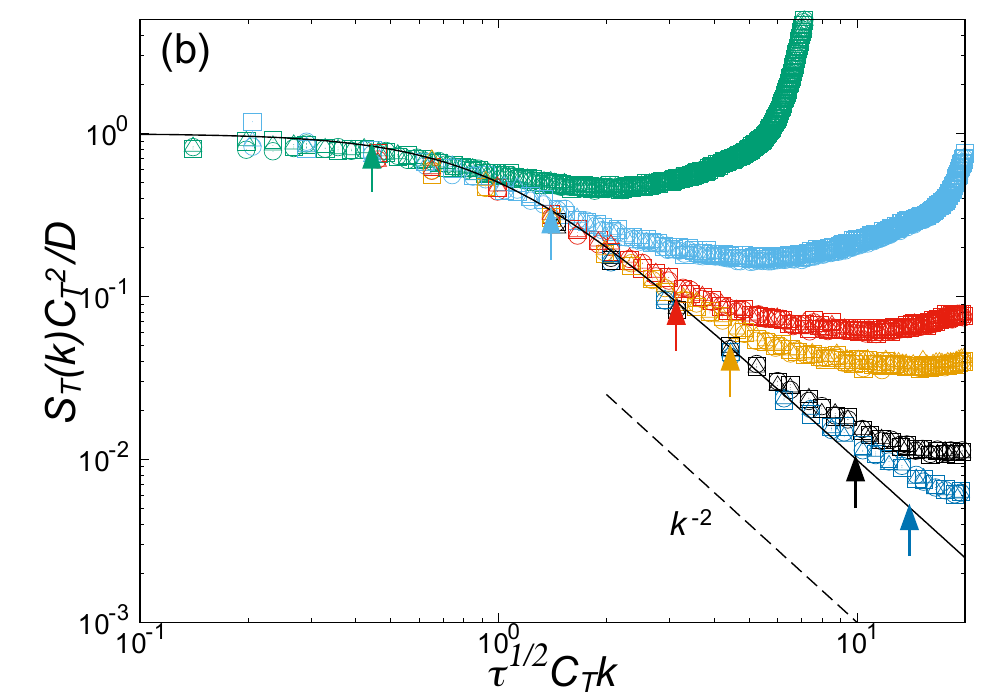}
\caption{
Correlation function of the transverse displacements.
(a) $S_T(k)$ is plotted as a function of wavenumber $k$ for various persistent times $\tau$; $\tau = 10^{-8}$~(purple), $10^{-1}$~(green), $10^0$~(cyan), $5 \times 10^0$~(red), $10^1$~(yellow), $5 \times 10^1$~(black), and $10^2$~(blue).
(b) $S_T(k) C_T^2 / D$ is plotted as a function of $\sqrt{\tau} C_T k$. 
The symbols represent simulation data; circles for AOUP glass, squares for ABP glass, and triangles for RTP glass.
The solid lines indicate the theoretical prediction of Eq.~(\ref{theory_Sk_scale}) with $\alpha =T$.
In (b), the vertical arrows indicate $\tau^{1/2} \omega_\text{BP}$, and the dashed line indicates $\propto k^{-2}$.
}
\label{fig.STk}
\end{figure}

\begin{figure}[t]
\centering
\includegraphics[width=42.5mm]{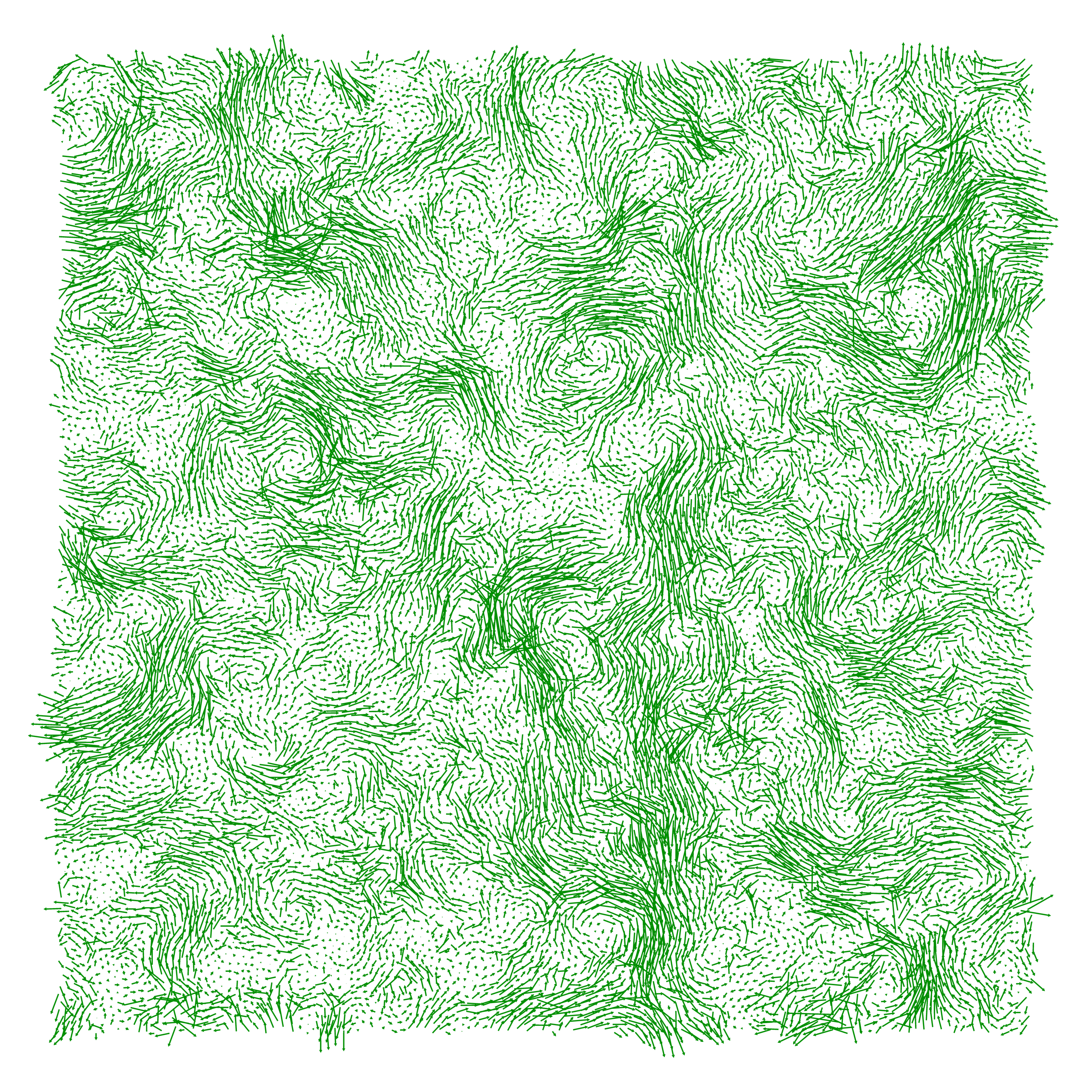}
\includegraphics[width=42.5mm]{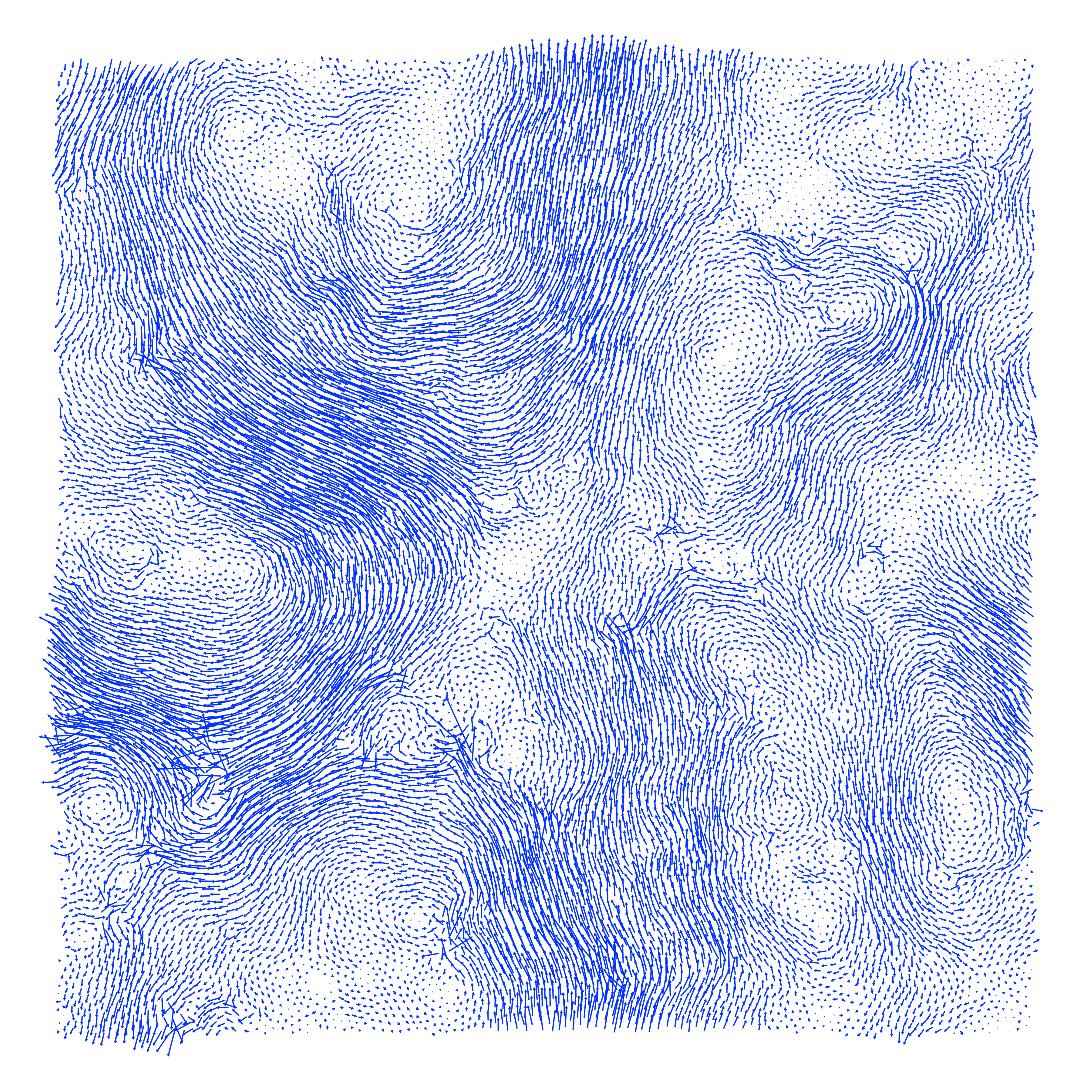}
\caption{
Displacement fields in the AOUP glass for $\tau=10^{-1}$~(left panel, green arrows) and for $\tau=10^{2}$~(right panel, blue arrows).
We plot $\vb*{u} = (\vec{u}_1, \vec{u}_2, \cdots, \vec{u}_N)$ using arrows.
To improve clarity, we multiply the arrow lengths by $50$ in the left panel and by $250$ in the right panel.
}
\label{fig.visual}
\end{figure}

We now discuss the spatial correlations of the longitudinal and transverse displacement fields, denoted as $S_L(k)$ and $S_T(k)$, respectively.
These correlation functions are calculated in our numerical simulations using Eq.~(\ref{def_sk}).
In Figs.~\ref{fig.SLk}(a) and \ref{fig.STk}(a), we present $S_L(k)$ and $S_T(k)$ for three different active glasses with varying persistent times $\tau$. 
When $\tau = 10^{-8}$, $S_L(k)$ and $S_T(k)$ exhibit a plateau in a wide range of $k$.
This is consistent with the known results for thermal glasses~\cite{Ikeda_2015}.
However, as $\tau$ is increased, both $S_L(k)$ and $S_T(k)$ become smaller at large $k$, indicating that short-wavelength excitations are strongly suppressed.
This indicates that the fluctuations of active glasses exhibit apparent collective motions as the long-wavelength excitations are emphasized~\cite{Henkes_2020}.
These apparent collective motions only appear when the persistence time $\tau$ is sufficiently large.
Notably, the above observations for all three active glasses are identical.

In parallel, we derived the theoretical formula for these correlation functions by substituting Eq.~(\ref{eq.corfl2}) into Eq.~(\ref{def_sk}), as
\begin{equation} \label{theory_Sk}
\begin{aligned}
& S_L(k) = \frac{Dk^2}{\gamma N} \sum_{i=1}^N \sum_{j=1}^N \sum_{l=1}^{2N-2} \frac{1}{\omega_l^2(\tau \omega_l^2 + 1)} \\
&\times \left( \hat{\vec{k}} \cdot \vec{e}_{l,i} \right) \left( \hat{\vec{k}} \cdot \vec{e}^{\dagger}_{l,j} \right) \exp\left[ i\left( \vec{k} \cdot \left( \vec{r}_{0,j} - \vec{r}_{0,i} \right) \right) \right],\\
& S_T(k) = \frac{Dk^2}{\gamma N} \sum_{i=1}^N \sum_{j=1}^N \sum_{l=1}^{2N-2} \frac{1}{\omega_l^2(\tau \omega_l^2 + 1)} \\
&\times \left[ \vec{e}_{l,i}  - \hat{\vec{k}} \left( \hat{\vec{k}} \cdot \vec{e}_{l,i} \right) \right] \left[ \vec{e}^{\dag}_{l,j} - \hat{\vec{k}} \left(\hat{\vec{k}} \cdot \vec{e}^{\dag}_{l,j} \right) \right] \\
&\times \exp\left[ i\left( \vec{k} \cdot \left( \vec{r}_{0,j} - \vec{r}_{0,i} \right) \right) \right].
\end{aligned}
\end{equation}
%
Here, we focus on the low-wavenumber regime in Eq.~(\ref{theory_Sk}).
We can use conventional elasticity theory to approximate the eigenmodes in this regime~\cite{Caprini2023}.
Specifically, in the two-dimensional continuum solid, we can approximate the dynamical matrix Eq.~(\ref{dmat}) as the operator $\displaystyle{\frac{1}{\gamma \rho} \left( K \nabla \nabla + G \nabla^2 \right)}$, which acts on the continuum displacement fields.
Here, $K$ and $G$ are the bulk and shear modulus, respectively, and $\rho = N/V$ is the number density.
We can solve the eigenvalue problem of this operator and obtain the eigenmodes specified by $l = \vec{k}\alpha$ where $\vec{k}$ is the wavevector and $\alpha$ is the polarization with $\alpha = L$ and $T$ denoting the longitudinal and the transverse modes, respectively.
The eigenfrequency can be expressed as
\begin{equation} \label{dispersion_relation}
\omega_{\vec{k}\alpha} = C_{\alpha} k,
\end{equation}
where $k=|\vec{k}|$ is wavenumber, and 
\begin{equation}
C_L = \sqrt{\frac{K+G}{\gamma \rho}},\qquad C_T = \sqrt{\frac{G}{\gamma \rho}}.
\end{equation}
$C_{\alpha}$ is similar to the sound speed, but its physical dimension is the length divided by the square root of time since that of the ``frequency" is the inverse of the square root of time~(see also the description of ``frequency" in Eq.~(\ref{def.frequency})).
The eigenvector is $\vb*{e}_{\vec{k}\alpha} = \left[ \vec{e}_{\vec{k}\alpha,1},\vec{e}_{\vec{k}\alpha,2},\cdots,\vec{e}_{\vec{k}\alpha,N} \right]$ with
\begin{equation} \label{continuum_eigenvec}
\vec{e}_{\vec{k}\alpha,i} = \frac{\vec{p}_{\alpha}}{\sqrt{N}} \exp\left( i \vec{k} \cdot \vec{r}_{0,i} \right),
\end{equation}
where $\vec{p}_{\alpha}$ is the polarization vector.
By plugging these Eqs.~(\ref{dispersion_relation}) and~(\ref{continuum_eigenvec}) into Eq.~(\ref{theory_Sk}), we arrive at
\begin{equation} \label{theory_Sk_scale}
S_\alpha (k) = \frac{D}{\gamma C_{\alpha}^2(\tau C_{\alpha}^2 k^2 + 1)}. 
\end{equation}
We note that this asymptotic law should only hold at low wavenumbers $k$ because modes above the boson peak frequency $\omega_\text{BP}$ are not plane-wave-like, where the linear dispersion relation in Eq.~\eqref{dispersion_relation} and the eigenvector in Eq.~(\ref{continuum_eigenvec}) are no longer applicable.

We can extract the characteristic length scale for the displacement fields from Eq.~(\ref{theory_Sk_scale}), as
\begin{equation} \label{Sk_length}
\xi_{\alpha} = \sqrt{\tau} C_\alpha.
\end{equation}
For $k \ll \xi_{\alpha}^{-1}$, Eq.~(\ref{theory_Sk_scale}) can be simplified to
\begin{equation} \label{theory_Sk_scale2}
S_{\alpha} (k) \simeq \frac{D}{\gamma C_{\alpha}^2}. 
\end{equation}
In the long-wavelength limit, $S_{\alpha} (k)$ becomes independent of $k$. 
Eq.~(\ref{theory_Sk_scale2}) is exactly the same as those for the thermal system, which is consistent with the equipartition law~\cite{Ikeda_2015}. 
On the other hand, for $k \gg \xi_{\alpha}^{-1}$, Eq.~(\ref{theory_Sk_scale}) can be simplified to
\begin{equation} \label{theory_Sk_scale3}
S_\alpha (k) \simeq \frac{D}{\gamma C_{\alpha}^2} \frac{1}{(\xi_{\alpha} k)^2}. 
\end{equation}
In the short-wavelength limit, $S_{\alpha} (k)$ decays as $k^{-2}$.
These results are reasonable based on our previous discussions.
We demonstrated that the displacements along the mode $l$ are strongly dampened when the quasi-static condition $\omega_l^2 \gg \tau^{-1}$ is met.
As the frequency can be converted into the wavenumber via $\omega = C_{\alpha} k$, we can translate the quasi-static condition into the wavenumber domain as $k \gg \tau^{-1/2} C_{\alpha}^{-1} = \xi_{\alpha}^{-1}$.
Therefore, $k \gg \xi_{\alpha}^{-1}$ is the quasi-static condition in the wavenumber domain.
In the quasi-static state, the displacements are significantly suppressed by an additional factor of $\omega_l^{-2}$, which rationalizes Eq.~(\ref{theory_Sk_scale3}).
Based on this discussion, we refer to $\xi_{\alpha}$ as the quasi-static length.
Interestingly, the quasi-static length increases as $\sqrt{\tau}$, which indicates that the longer the active forces persist, the more the wavelength excitations become quasi-static and dampened.

From Eq.~(\ref{Sk_length}), we can describe the quasi-static length $\xi_{\alpha}$ as
\begin{equation} \label{lengthscale}
\xi_L = \sqrt{\frac{(K + G) \tau}{\gamma \rho}}, \qquad
\xi_T = \sqrt{\frac{G \tau}{\gamma \rho}}.
\end{equation}
These lengths can also be estimated in a different way.
When we consider a transverse wave with the wavenumber $k$, the restoring force acting on this wave (per particle) in continuum mechanics is estimated as $Gk^2/\rho$, and its relaxation time is $\gamma \rho/(G k^2)$.
If this relaxation time becomes longer than the persistence time $\tau$, the quasi-static nature emerges.
This gives an estimate of the characteristic wavenumber as $k_T \sim \sqrt{\gamma \rho/(G \tau)}$, which reproduces $\xi_T$ in Eq.~(\ref{lengthscale}). 
Likewise, for a longitudinal wave, we estimate the characteristic wavenumber as $k_L \sim \sqrt{\gamma \rho/[(K + G) \tau]}$, which reproduces $\xi_L$ in Eq.~(\ref{lengthscale}). 

In order to verify the asymptotic behavior given by Eq.~(\ref{theory_Sk_scale}), we plotted the simulation data in the form of $S_{\alpha}(k) \gamma C_{\alpha}^2 /D$ against $\xi_{\alpha} k$ in Figs.~\ref{fig.SLk}(b) and \ref{fig.STk}(b).
These data show consistency with the asymptotic law at low $k$.
However, at high $k$, the data deviates from the asymptotic law as expected from the fact that the dispersion relation Eq.~\eqref{dispersion_relation} and the eigenvector in Eq.~(\ref{continuum_eigenvec}) are not valid above the boson peak frequency $\omega_\text{BP}$.
To discuss this point further, let us define the characteristic length scale associated with the boson peak as $\xi_\text{BP} = C_\alpha / \omega_\text{BP}$.
The arrows in Figs.~\ref{fig.SLk}(b) and \ref{fig.STk}(b) indicate $\tau^{1/2} C_{\alpha} / \xi_\text{BP} = \tau^{1/2} \omega_\text{BP}$, which confirms that the deviation actually appears at $k \simeq \xi_\text{BP}^{-1}$.

The collective motions in active glasses are more noticeable as the large-wavenumber~(or short-wavelength) excitations are more strongly suppressed.
This is apparent by comparing the data for $\tau = 10^{-1}$~(green symbols) and those for $\tau = 10^2$~(blue symbols) in Figs.~\ref{fig.SLk}(b) and~\ref{fig.STk}(b).
We also visualize the displacement fields for these two cases in Fig.~\ref{fig.visual}.
For $\tau = 10^{-1}$, the structure factor does not show a strong $k$ dependence, which means that the collective motions are less visible.
In contrast, for $\tau = 10^2$, the structure factor displays a strong $k$ dependence with significant suppressions at large $k$, thus resulting in the emergence of collective motions.
To observe the collective motions, the condition $\tau^{1/2} C_{\alpha} / \xi_\text{BP} \gg 1$ is required, which is equivalent to the condition that the quasi-static length $\xi_\alpha$ is longer than $\xi_\text{BP}$, \textit{i.e.}, $\xi_\alpha > \xi_\text{BP}$.
This condition can be transformed into
\begin{eqnarray}\label{eqbp}
\tau  \gg \omega_\text{BP}^{-2}. 
\end{eqnarray}
In other words, the apparent collective motions in active glasses emerge when the persistence time of active force is longer than the characteristic relaxation time of the modes at the boson peak.

\subsection{Non-Gaussian parameter}
%
\begin{figure}[t]
\centering
\includegraphics[width=80mm]{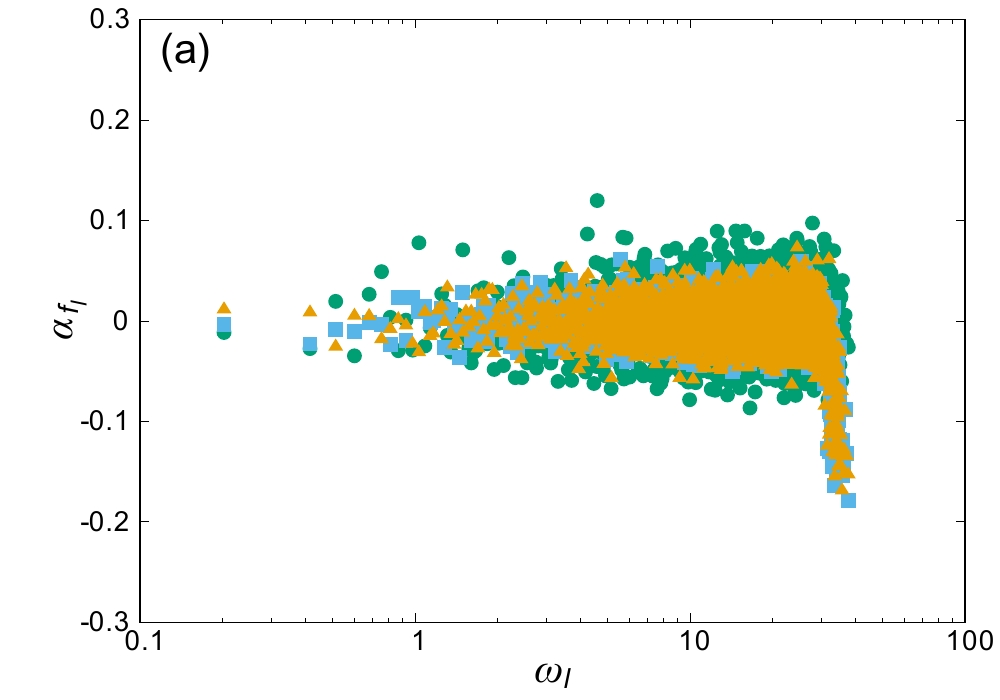}
\includegraphics[width=80mm]{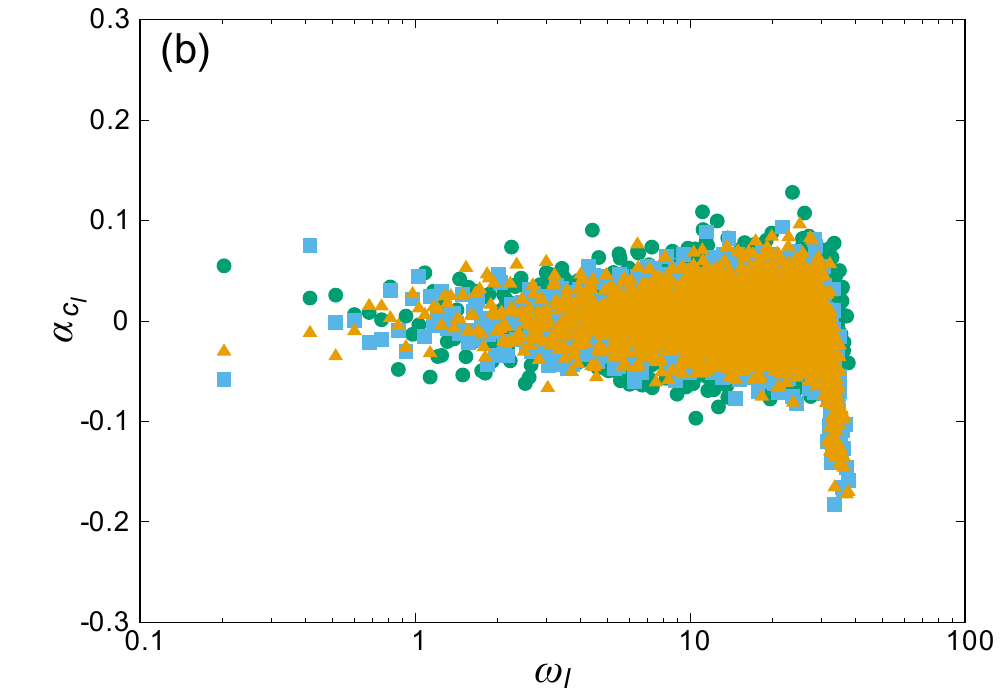}
\caption{
(a) NGP $\alpha_{f_l}$ of $P(f_l)$ and (b) NGP $\alpha_{c_l}$ of $P(c_l)$ are plotted as a function of $\omega_l$. 
The persistent time is $\tau = 10^0$.
The symbols represent simulation data: AOUP glass~(green circles), ABP glass~(cyan squares), and RTP glass~(yellow triangles).
} \label{fig.NGP}
\end{figure}

We finally discuss the non-Gaussianity of $f_l$ and $c_l$ using the NGPs, $\alpha_{f_l}$ and $\alpha_{c_l}$.
We calculate $\alpha_{f_l}$ and $\alpha_{c_l}$ through numerical simulations, using Eqs.~(\ref{def_fNGP}) and (\ref{def_cNGP}), and present the data in Fig.~\ref{fig.NGP}.
It is important to note that while the figure is for $\tau = 10^0$, we have confirmed similar results for $\tau = 10^{-1}, 10^1, 10^2$.

We begin by examining the non-Gaussianity of $f_l$.
For the AOUP glass, we observe that $\alpha_{f_l}$ is approximately zero across all modes, which is in line with our discussion in Sec.~\ref{sec.activenoises.aoup}.
For the ABP and RTP glasses, we observe that $\alpha_{f_l}$ is approximately zero in the frequency range $\omega_l \lesssim 30$.
As illustrated in Fig~\ref{fig.vib}(b) in Sec.~\ref{sec.Numericalmodel}, the eigenmodes in this frequency range are spatially extended.
Consequently, this outcome is consistent with our Sec.~\ref{sec.activenoises.aoup}.
Namely, while the active forces acting on a single particle are highly non-Gaussian, the central limit theorem indicates $\alpha_{f_l} = 0$ for a spatially extended mode $l$.
Instead, we observe the deviation of $\alpha_{f_l}$ from zero in $\omega_l \gtrsim 30$, where the eigenmodes are spatially localized~(see Fig~\ref{fig.vib}(b)).
Specifically, we find that $\alpha_{f_l}$ is negative, which implies that the probability distribution $P(f_l)$ shows a weakly decaying tail.
We contrast this result with an idealized case of a single RTP in one dimension, where the distribution of the active force $f$ is the sum of two delta peaks, given by
\begin{equation}
P(f) = \frac{1}{2}\delta(f - \gamma v_0) + \frac{1}{2}\delta(f + \gamma v_0).
\end{equation}
The NGP for this distribution is $\alpha_f = -2/3$.
We thus attribute the negative $\alpha_{f_l}$ for the localized mode in the ABP and RTP glasses to a remnant of this idealized scenario, even though the observed $\alpha_{f_l}$ is still larger than $-2/3$, which may reflect a slight collective nature of these localized modes.

Next, we focus on the non-Gaussianity of $c_l$.
It is observed that $\alpha_{c_l}$ in Fig.~\ref{fig.NGP}(b) is very similar to $\alpha_{f_l}$ in Fig.~\ref{fig.NGP}(a).
This is expected as $c_l$ is a linear combination of $f_l$ as seen in Eq.~(\ref{theory_c_l}), and the linear combination of correlated Gaussian variables is still a Gaussian variable~\cite{Lemons_2003}.
Hence, it is natural for $\alpha_{c_l} \approx 0$ when $\alpha_{f_l} \approx 0$.
On the other hand, the deviation $\alpha_{c_l} < 0$ in the high-frequency regime of $\omega_l \gtrsim 30$ in the ABP and RTP glasses can be understood in the light of $\alpha_{f_l} < 0$.
In previous studies~\cite{Malakar_2020, Dhar_2019} in a single RTP in a one-dimensional trap and a single ABP in a two-dimensional trap, it was shown that non-Gaussianity of the active force causes the accumulation of particles near the wall, resulting in a particle distribution with two sharp peaks.
We attribute the negative $\alpha_{c_l}$ to a remnant of such idealized situations.

We thus conclude that the displacements along all the modes in the AOUP glass follow a Gaussian distribution.
The same holds for spatially extended modes in the ABP and RTP glasses.
However, we observe a subtle deviation in the localized mode in the ABP and RTP glasses.
This deviation could be a vestige of the non-Gaussianity of a single ABP and RTP dynamics in a trap.

\section{Conclusion}
In this paper, we conducted both numerical and theoretical analyses to investigate structural fluctuations in the active glasses.
We developed a general formalism to study the properties of solid-state active glasses and applied it to three different models: the AOUP glass, the ABP glass, and the RTP glass.
This formalism allowed us to express various observables such as the average excitation energies of modes $E_l$, the MSD $R(t)$, and the structure factors of longitudinal and transverse displacement fields, $S_L(k)$ and $S_T(k)$, in terms of normal modes and two-time correlations of active forces.
Our findings showed that the two-time correlations in the AOUP, ABP, and RTP glasses are equivalent after proper reparametrization, and thus, the listed observables are also identical among the three models.

We used formalism to numerically evaluate the observables for a typical configuration of active glasses.
Our key observation is that the active forces behave like random noise for the low-frequency modes, while they behave like constant forces for the high-frequency modes, as described in Eqs.~(\ref{eq.corfl2}) to~(\ref{eq.corfl4}).
As a result, the low-frequency modes are excited, similar to thermal systems, while the dynamics along high-frequency modes are quasi-static, resulting in strong suppression of excitations.
We then found that this mechanism dominates the observables $E_l$, $R(t)$, and $S_L(k),\ S_T(k)$ in active glasses. 
The average excitation energy $E_l$ follows the equipartition law very well for the low-frequency modes, while it sharply drops at the high-frequency modes, as shown in Eqs.~(\ref{theory_E}) to~(\ref{theory_E3}).
Accordingly, the MSD $R(t)$ is highly suppressed in both short and long time regimes.
Notably, the effective velocity of each particle in active glasses becomes much smaller than that for non-interacting active particles, as shown in Eq.~\eqref{theory_MSD2}. This is because the dynamics along the high-frequency modes become quasi-static, and they cannot contribute to the particles' displacements. 
Furthermore, the structure factors of the longitudinal and transverse displacement fields, denoted by $S_L(k)$ and $S_T(k)$ respectively, exhibit collective behaviors as described in Eqs.~\eqref{theory_Sk_scale} to \eqref{theory_Sk_scale3}.
Specifically, short-wavelength excitations are strongly suppressed due to the quasi-static nature of these modes, while long-wavelength excitations are emphasized.
We demonstrated that the characteristic length scales of these collective behaviors are determined by the persistence time of the active forces and the elastic moduli of active glasses, as shown in Eq.~\eqref{lengthscale}.
It is important to note that these collective motions are only observable when the persistence time is longer than the relaxation time of the modes at the boson peak, as presented in Eq.~\eqref{eqbp}.

We also conducted measurements on the non-Gaussian parameters of the active forces and displacements along the vibrational modes to better understand the fluctuations.
In the case of the AOUP glass, we observed that all non-Gaussian parameters are zero, indicating that the active forces follow Gaussian statistics exactly.
Interestingly, we also found that for the ABP and RTP glasses, most of the vibrational modes have zero non-Gaussian parameters, despite the active forces following non-Gaussian statistics.
We attribute this emergence of Gaussianity to the spatially extended nature of the vibrational modes, with non-Gaussianity only persisting in spatially localized modes.

In this work, we conducted a comprehensive analysis of the dynamics of active glasses using harmonic description.
Previously, Henke \textit{et al.}~\cite{Henkes_2020} have studied the velocity fields of ABP glasses using harmonic description to analyze confluent cells.
They obtained Eq.~(\ref{theory_E}) for excitational energy, similar equation for Eq.~(\ref{theory_MSD2}) for mean square average of velocity.
They also obtained Eq.~(\ref{lengthscale}) for correlation length of velocity fields although our Eq.~(\ref{lengthscale}) is for displacement fields.  
Our work builds on this by considering three different active glass models, analyzing the displacement fields, and discussing the impact of the characteristic normal modes in glasses.
These results will be useful for analyzing the displacement fields in a wide range of active glasses.

Towards the end of this paper, we would like to suggest a possible extension of our work.
Firstly, we find it interesting to explore the structural fluctuations in active glasses with lower densities.
The present work focused on densities above the jamming point, where each particle is always in contact with other particles.
However, when the density is below the jamming point, the system remains in an active glass state while each particle is caged due to frequent collisions with other particles, as in the case of hard-sphere glasses~\cite{Pusey1986,Brito2009,Henkes_2012}.
The dynamics of such active glasses are highly complex, and the harmonic description cannot be applied directly.
Therefore, studying the structural fluctuations of such lower-density active glasses would be an important and fascinating topic.
Secondly, we suggest studying the structural relaxations of active glasses.
In thermal glasses, the structural relaxation has been discussed in terms of the excitation of quasi-localized vibrational modes~\cite{Widmer_2008}.
It is highly non-trivial to understand how the active forces excite these modes and trigger the rearrangement of particles.
The present work provides a solid foundation to study such non-linear dynamics of active glasses.

\section*{Acknowledgments}
We express our gratitude to Yuta Kuroda for providing valuable suggestions and engaging in productive discussions with us.
This work was supported by JSPS KAKENHI Grant Numbers 18H05225, 19H01812, 20H00128, 20H01868, 22K03543, 23H04495.

\section*{Author declarations}
The authors declare no conflicts of interest.

\section*{Data availability}
The data that support the findings of this study are available within the article.

\bibliographystyle{apsrev4-2}
\bibliography{reference}

\begin{thebibliography}{61}%
\makeatletter
\providecommand \@ifxundefined [1]{%
 \@ifx{#1\undefined}
}%
\providecommand \@ifnum [1]{%
 \ifnum #1\expandafter \@firstoftwo
 \else \expandafter \@secondoftwo
 \fi
}%
\providecommand \@ifx [1]{%
 \ifx #1\expandafter \@firstoftwo
 \else \expandafter \@secondoftwo
 \fi
}%
\providecommand \natexlab [1]{#1}%
\providecommand \enquote  [1]{``#1''}%
\providecommand \bibnamefont  [1]{#1}%
\providecommand \bibfnamefont [1]{#1}%
\providecommand \citenamefont [1]{#1}%
\providecommand \href@noop [0]{\@secondoftwo}%
\providecommand \href [0]{\begingroup \@sanitize@url \@href}%
\providecommand \@href[1]{\@@startlink{#1}\@@href}%
\providecommand \@@href[1]{\endgroup#1\@@endlink}%
\providecommand \@sanitize@url [0]{\catcode `\\12\catcode `\$12\catcode
  `\&12\catcode `\#12\catcode `\^12\catcode `\_12\catcode `\%12\relax}%
\providecommand \@@startlink[1]{}%
\providecommand \@@endlink[0]{}%
\providecommand \url  [0]{\begingroup\@sanitize@url \@url }%
\providecommand \@url [1]{\endgroup\@href {#1}{\urlprefix }}%
\providecommand \urlprefix  [0]{URL }%
\providecommand \Eprint [0]{\href }%
\providecommand \doibase [0]{https://doi.org/}%
\providecommand \selectlanguage [0]{\@gobble}%
\providecommand \bibinfo  [0]{\@secondoftwo}%
\providecommand \bibfield  [0]{\@secondoftwo}%
\providecommand \translation [1]{[#1]}%
\providecommand \BibitemOpen [0]{}%
\providecommand \bibitemStop [0]{}%
\providecommand \bibitemNoStop [0]{.\EOS\space}%
\providecommand \EOS [0]{\spacefactor3000\relax}%
\providecommand \BibitemShut  [1]{\csname bibitem#1\endcsname}%
\let\auto@bib@innerbib\@empty
\bibitem [{\citenamefont {Marchetti}\ \emph {et~al.}(2013)\citenamefont
  {Marchetti}, \citenamefont {Joanny}, \citenamefont {Ramaswamy}, \citenamefont
  {Liverpool}, \citenamefont {Prost}, \citenamefont {Rao},\ and\ \citenamefont
  {Simha}}]{Marchetti2013}%
  \BibitemOpen
  \bibfield  {author} {\bibinfo {author} {\bibfnamefont {M.~C.}\ \bibnamefont
  {Marchetti}}, \bibinfo {author} {\bibfnamefont {J.~F.}\ \bibnamefont
  {Joanny}}, \bibinfo {author} {\bibfnamefont {S.}~\bibnamefont {Ramaswamy}},
  \bibinfo {author} {\bibfnamefont {T.~B.}\ \bibnamefont {Liverpool}}, \bibinfo
  {author} {\bibfnamefont {J.}~\bibnamefont {Prost}}, \bibinfo {author}
  {\bibfnamefont {M.}~\bibnamefont {Rao}},\ and\ \bibinfo {author}
  {\bibfnamefont {R.~A.}\ \bibnamefont {Simha}},\ }\href
  {https://doi.org/10.1103/RevModPhys.85.1143} {\bibfield  {journal} {\bibinfo
  {journal} {Rev. Mod. Phys.}\ }\textbf {\bibinfo {volume} {85}},\ \bibinfo
  {pages} {1143} (\bibinfo {year} {2013})}\BibitemShut {NoStop}%
\bibitem [{\citenamefont {Vicsek}\ \emph {et~al.}(1995)\citenamefont {Vicsek},
  \citenamefont {Czir\'ok}, \citenamefont {Ben-Jacob}, \citenamefont {Cohen},\
  and\ \citenamefont {Shochet}}]{Vicsek_1995}%
  \BibitemOpen
  \bibfield  {author} {\bibinfo {author} {\bibfnamefont {T.}~\bibnamefont
  {Vicsek}}, \bibinfo {author} {\bibfnamefont {A.}~\bibnamefont {Czir\'ok}},
  \bibinfo {author} {\bibfnamefont {E.}~\bibnamefont {Ben-Jacob}}, \bibinfo
  {author} {\bibfnamefont {I.}~\bibnamefont {Cohen}},\ and\ \bibinfo {author}
  {\bibfnamefont {O.}~\bibnamefont {Shochet}},\ }\href
  {https://doi.org/10.1103/PhysRevLett.75.1226} {\bibfield  {journal} {\bibinfo
   {journal} {Phys. Rev. Lett.}\ }\textbf {\bibinfo {volume} {75}},\ \bibinfo
  {pages} {1226} (\bibinfo {year} {1995})}\BibitemShut {NoStop}%
\bibitem [{\citenamefont {Cates}\ and\ \citenamefont
  {Tailleur}(2015)}]{Cates2015}%
  \BibitemOpen
  \bibfield  {author} {\bibinfo {author} {\bibfnamefont {M.~E.}\ \bibnamefont
  {Cates}}\ and\ \bibinfo {author} {\bibfnamefont {J.}~\bibnamefont
  {Tailleur}},\ }\href
  {https://doi.org/https://doi.org/10.1146/annurev-conmatphys-031214-014710}
  {\bibfield  {journal} {\bibinfo  {journal} {Annual Review of Condensed Matter
  Physics}\ }\textbf {\bibinfo {volume} {6}},\ \bibinfo {pages} {219} (\bibinfo
  {year} {2015})}\BibitemShut {NoStop}%
\bibitem [{\citenamefont {Alert}\ \emph {et~al.}(2022)\citenamefont {Alert},
  \citenamefont {Casademunt},\ and\ \citenamefont {Joanny}}]{Alert2022}%
  \BibitemOpen
  \bibfield  {author} {\bibinfo {author} {\bibfnamefont {R.}~\bibnamefont
  {Alert}}, \bibinfo {author} {\bibfnamefont {J.}~\bibnamefont {Casademunt}},\
  and\ \bibinfo {author} {\bibfnamefont {J.-F.}\ \bibnamefont {Joanny}},\
  }\href
  {https://doi.org/https://doi.org/10.1146/annurev-conmatphys-082321-035957}
  {\bibfield  {journal} {\bibinfo  {journal} {Annual Review of Condensed Matter
  Physics}\ }\textbf {\bibinfo {volume} {13}},\ \bibinfo {pages} {143}
  (\bibinfo {year} {2022})}\BibitemShut {NoStop}%
\bibitem [{\citenamefont {Ramaswamy}\ \emph {et~al.}(2003)\citenamefont
  {Ramaswamy}, \citenamefont {Simha},\ and\ \citenamefont
  {Toner}}]{Ramaswamy_2003}%
  \BibitemOpen
  \bibfield  {author} {\bibinfo {author} {\bibfnamefont {S.}~\bibnamefont
  {Ramaswamy}}, \bibinfo {author} {\bibfnamefont {R.~A.}\ \bibnamefont
  {Simha}},\ and\ \bibinfo {author} {\bibfnamefont {J.}~\bibnamefont {Toner}},\
  }\href {https://doi.org/10.1209/epl/i2003-00346-7} {\bibfield  {journal}
  {\bibinfo  {journal} {Europhysics Letters}\ }\textbf {\bibinfo {volume}
  {62}},\ \bibinfo {pages} {196} (\bibinfo {year} {2003})}\BibitemShut
  {NoStop}%
\bibitem [{\citenamefont {Angelini}\ \emph {et~al.}(2011)\citenamefont
  {Angelini}, \citenamefont {Hannezo}, \citenamefont {Trepat}, \citenamefont
  {Marquez}, \citenamefont {Fredberg},\ and\ \citenamefont
  {Weitz}}]{Angelini_2011}%
  \BibitemOpen
  \bibfield  {author} {\bibinfo {author} {\bibfnamefont {T.~E.}\ \bibnamefont
  {Angelini}}, \bibinfo {author} {\bibfnamefont {E.}~\bibnamefont {Hannezo}},
  \bibinfo {author} {\bibfnamefont {X.}~\bibnamefont {Trepat}}, \bibinfo
  {author} {\bibfnamefont {M.}~\bibnamefont {Marquez}}, \bibinfo {author}
  {\bibfnamefont {J.~J.}\ \bibnamefont {Fredberg}},\ and\ \bibinfo {author}
  {\bibfnamefont {D.~A.}\ \bibnamefont {Weitz}},\ }\href@noop {} {\bibfield
  {journal} {\bibinfo  {journal} {Proceedings of the National Academy of
  Sciences}\ }\textbf {\bibinfo {volume} {108}},\ \bibinfo {pages} {4714}
  (\bibinfo {year} {2011})}\BibitemShut {NoStop}%
\bibitem [{\citenamefont {Berthier}\ \emph {et~al.}(2019)\citenamefont
  {Berthier}, \citenamefont {Flenner},\ and\ \citenamefont
  {Szamel}}]{Berthier2019}%
  \BibitemOpen
  \bibfield  {author} {\bibinfo {author} {\bibfnamefont {L.}~\bibnamefont
  {Berthier}}, \bibinfo {author} {\bibfnamefont {E.}~\bibnamefont {Flenner}},\
  and\ \bibinfo {author} {\bibfnamefont {G.}~\bibnamefont {Szamel}},\ }\href
  {https://doi.org/10.1063/1.5093240} {\bibfield  {journal} {\bibinfo
  {journal} {The Journal of Chemical Physics}\ }\textbf {\bibinfo {volume}
  {150}},\ \bibinfo {pages} {200901} (\bibinfo {year} {2019})}\BibitemShut
  {NoStop}%
\bibitem [{\citenamefont {Janssen}(2019)}]{Janssen_2019}%
  \BibitemOpen
  \bibfield  {author} {\bibinfo {author} {\bibfnamefont {L.~M.~C.}\
  \bibnamefont {Janssen}},\ }\href {https://doi.org/10.1088/1361-648X/ab3e90}
  {\bibfield  {journal} {\bibinfo  {journal} {Journal of Physics: Condensed
  Matter}\ }\textbf {\bibinfo {volume} {31}},\ \bibinfo {pages} {503002}
  (\bibinfo {year} {2019})}\BibitemShut {NoStop}%
\bibitem [{\citenamefont {Zhou}\ \emph {et~al.}(2009)\citenamefont {Zhou},
  \citenamefont {Trepat}, \citenamefont {Park}, \citenamefont {Lenormand},
  \citenamefont {Oliver}, \citenamefont {Mijailovich}, \citenamefont {Hardin},
  \citenamefont {Weitz}, \citenamefont {Butler},\ and\ \citenamefont
  {Fredberg}}]{Zhou_2009}%
  \BibitemOpen
  \bibfield  {author} {\bibinfo {author} {\bibfnamefont {E.}~\bibnamefont
  {Zhou}}, \bibinfo {author} {\bibfnamefont {X.}~\bibnamefont {Trepat}},
  \bibinfo {author} {\bibfnamefont {C.}~\bibnamefont {Park}}, \bibinfo {author}
  {\bibfnamefont {G.}~\bibnamefont {Lenormand}}, \bibinfo {author}
  {\bibfnamefont {M.}~\bibnamefont {Oliver}}, \bibinfo {author} {\bibfnamefont
  {S.}~\bibnamefont {Mijailovich}}, \bibinfo {author} {\bibfnamefont
  {C.}~\bibnamefont {Hardin}}, \bibinfo {author} {\bibfnamefont
  {D.}~\bibnamefont {Weitz}}, \bibinfo {author} {\bibfnamefont
  {J.}~\bibnamefont {Butler}},\ and\ \bibinfo {author} {\bibfnamefont
  {J.}~\bibnamefont {Fredberg}},\ }\href@noop {} {\bibfield  {journal}
  {\bibinfo  {journal} {Proceedings of the National Academy of Sciences}\
  }\textbf {\bibinfo {volume} {106}},\ \bibinfo {pages} {10632} (\bibinfo
  {year} {2009})}\BibitemShut {NoStop}%
\bibitem [{\citenamefont {Nnetu}\ \emph {et~al.}(2012)\citenamefont {Nnetu},
  \citenamefont {Knorr}, \citenamefont {Kas},\ and\ \citenamefont
  {Zink}}]{Nnetu_2012}%
  \BibitemOpen
  \bibfield  {author} {\bibinfo {author} {\bibfnamefont {K.~D.}\ \bibnamefont
  {Nnetu}}, \bibinfo {author} {\bibfnamefont {M.}~\bibnamefont {Knorr}},
  \bibinfo {author} {\bibfnamefont {J.}~\bibnamefont {Kas}},\ and\ \bibinfo
  {author} {\bibfnamefont {M.}~\bibnamefont {Zink}},\ }\href
  {https://doi.org/10.1088/1367-2630/14/11/115012} {\bibfield  {journal}
  {\bibinfo  {journal} {New Journal of Physics}\ }\textbf {\bibinfo {volume}
  {14}},\ \bibinfo {pages} {115012} (\bibinfo {year} {2012})}\BibitemShut
  {NoStop}%
\bibitem [{\citenamefont {Schoetz}\ \emph {et~al.}(2013)\citenamefont
  {Schoetz}, \citenamefont {Lanio}, \citenamefont {Talbot},\ and\ \citenamefont
  {Manning}}]{Schoetz_2013}%
  \BibitemOpen
  \bibfield  {author} {\bibinfo {author} {\bibfnamefont {E.-M.}\ \bibnamefont
  {Schoetz}}, \bibinfo {author} {\bibfnamefont {M.}~\bibnamefont {Lanio}},
  \bibinfo {author} {\bibfnamefont {J.~A.}\ \bibnamefont {Talbot}},\ and\
  \bibinfo {author} {\bibfnamefont {M.~L.}\ \bibnamefont {Manning}},\
  }\href@noop {} {\bibfield  {journal} {\bibinfo  {journal} {Journal of The
  Royal Society Interface}\ }\textbf {\bibinfo {volume} {10}},\ \bibinfo
  {pages} {20130726} (\bibinfo {year} {2013})}\BibitemShut {NoStop}%
\bibitem [{\citenamefont {Haeger}\ \emph {et~al.}(2014)\citenamefont {Haeger},
  \citenamefont {Krause}, \citenamefont {Wolf},\ and\ \citenamefont
  {Friedl}}]{Haeger2014}%
  \BibitemOpen
  \bibfield  {author} {\bibinfo {author} {\bibfnamefont {A.}~\bibnamefont
  {Haeger}}, \bibinfo {author} {\bibfnamefont {M.}~\bibnamefont {Krause}},
  \bibinfo {author} {\bibfnamefont {K.}~\bibnamefont {Wolf}},\ and\ \bibinfo
  {author} {\bibfnamefont {P.}~\bibnamefont {Friedl}},\ }\href
  {https://doi.org/https://doi.org/10.1016/j.bbagen.2014.03.020} {\bibfield
  {journal} {\bibinfo  {journal} {Biochimica et Biophysica Acta (BBA) - General
  Subjects}\ }\textbf {\bibinfo {volume} {1840}},\ \bibinfo {pages} {2386}
  (\bibinfo {year} {2014})},\ \bibinfo {note} {matrix-mediated cell behaviour
  and properties}\BibitemShut {NoStop}%
\bibitem [{\citenamefont {Garcia}\ \emph {et~al.}(2015)\citenamefont {Garcia},
  \citenamefont {Hannezo}, \citenamefont {Elgeti}, \citenamefont {Joanny},
  \citenamefont {Silberzan},\ and\ \citenamefont {Gov}}]{Garcia_2015}%
  \BibitemOpen
  \bibfield  {author} {\bibinfo {author} {\bibfnamefont {S.}~\bibnamefont
  {Garcia}}, \bibinfo {author} {\bibfnamefont {E.}~\bibnamefont {Hannezo}},
  \bibinfo {author} {\bibfnamefont {J.}~\bibnamefont {Elgeti}}, \bibinfo
  {author} {\bibfnamefont {J.-F.}\ \bibnamefont {Joanny}}, \bibinfo {author}
  {\bibfnamefont {P.}~\bibnamefont {Silberzan}},\ and\ \bibinfo {author}
  {\bibfnamefont {N.~S.}\ \bibnamefont {Gov}},\ }\href@noop {} {\bibfield
  {journal} {\bibinfo  {journal} {Proceedings of the National Academy of
  Sciences}\ }\textbf {\bibinfo {volume} {112}},\ \bibinfo {pages} {15314}
  (\bibinfo {year} {2015})}\BibitemShut {NoStop}%
\bibitem [{\citenamefont {Park}\ \emph {et~al.}(2016)\citenamefont {Park},
  \citenamefont {Atia}, \citenamefont {Mitchel}, \citenamefont {Fredberg},\
  and\ \citenamefont {Butler}}]{Park_2016}%
  \BibitemOpen
  \bibfield  {author} {\bibinfo {author} {\bibfnamefont {J.-A.}\ \bibnamefont
  {Park}}, \bibinfo {author} {\bibfnamefont {L.}~\bibnamefont {Atia}}, \bibinfo
  {author} {\bibfnamefont {J.~A.}\ \bibnamefont {Mitchel}}, \bibinfo {author}
  {\bibfnamefont {J.~J.}\ \bibnamefont {Fredberg}},\ and\ \bibinfo {author}
  {\bibfnamefont {J.~P.}\ \bibnamefont {Butler}},\ }\href@noop {} {\bibfield
  {journal} {\bibinfo  {journal} {Journal of cell science}\ }\textbf {\bibinfo
  {volume} {129}},\ \bibinfo {pages} {3375} (\bibinfo {year}
  {2016})}\BibitemShut {NoStop}%
\bibitem [{\citenamefont {Vishwakarma}\ \emph {et~al.}(2020)\citenamefont
  {Vishwakarma}, \citenamefont {Thurakkal}, \citenamefont {Spatz},\ and\
  \citenamefont {Das}}]{Vishwakarma_2020}%
  \BibitemOpen
  \bibfield  {author} {\bibinfo {author} {\bibfnamefont {M.}~\bibnamefont
  {Vishwakarma}}, \bibinfo {author} {\bibfnamefont {B.}~\bibnamefont
  {Thurakkal}}, \bibinfo {author} {\bibfnamefont {J.~P.}\ \bibnamefont
  {Spatz}},\ and\ \bibinfo {author} {\bibfnamefont {T.}~\bibnamefont {Das}},\
  }\href@noop {} {\bibfield  {journal} {\bibinfo  {journal} {Philosophical
  Transactions of the Royal Society B}\ }\textbf {\bibinfo {volume} {375}},\
  \bibinfo {pages} {20190391} (\bibinfo {year} {2020})}\BibitemShut {NoStop}%
\bibitem [{\citenamefont {Parry}\ \emph {et~al.}(2014)\citenamefont {Parry},
  \citenamefont {Surovtsev}, \citenamefont {Cabeen}, \citenamefont {O'Hern},
  \citenamefont {Dufresne},\ and\ \citenamefont {Jacobs-Wagner}}]{Parry2014}%
  \BibitemOpen
  \bibfield  {author} {\bibinfo {author} {\bibfnamefont {B.~R.}\ \bibnamefont
  {Parry}}, \bibinfo {author} {\bibfnamefont {I.~V.}\ \bibnamefont
  {Surovtsev}}, \bibinfo {author} {\bibfnamefont {M.~T.}\ \bibnamefont
  {Cabeen}}, \bibinfo {author} {\bibfnamefont {C.~S.}\ \bibnamefont {O'Hern}},
  \bibinfo {author} {\bibfnamefont {E.~R.}\ \bibnamefont {Dufresne}},\ and\
  \bibinfo {author} {\bibfnamefont {C.}~\bibnamefont {Jacobs-Wagner}},\
  }\href@noop {} {\bibfield  {journal} {\bibinfo  {journal} {Cell}\ }\textbf
  {\bibinfo {volume} {156}},\ \bibinfo {pages} {183} (\bibinfo {year}
  {2014})}\BibitemShut {NoStop}%
\bibitem [{\citenamefont {Nishizawa}\ \emph {et~al.}(2017)\citenamefont
  {Nishizawa}, \citenamefont {Fujiwara}, \citenamefont {Ikenaga}, \citenamefont
  {Nakajo}, \citenamefont {Yanagisawa},\ and\ \citenamefont
  {Mizuno}}]{Nishizawa2017}%
  \BibitemOpen
  \bibfield  {author} {\bibinfo {author} {\bibfnamefont {K.}~\bibnamefont
  {Nishizawa}}, \bibinfo {author} {\bibfnamefont {K.}~\bibnamefont {Fujiwara}},
  \bibinfo {author} {\bibfnamefont {M.}~\bibnamefont {Ikenaga}}, \bibinfo
  {author} {\bibfnamefont {N.}~\bibnamefont {Nakajo}}, \bibinfo {author}
  {\bibfnamefont {M.}~\bibnamefont {Yanagisawa}},\ and\ \bibinfo {author}
  {\bibfnamefont {D.}~\bibnamefont {Mizuno}},\ }\bibfield  {journal} {\bibinfo
  {journal} {Sci. Rep.}\ }\href {https://doi.org/10.1038/s41598-017-14883-y}
  {10.1038/s41598-017-14883-y} (\bibinfo {year} {2017})\BibitemShut {NoStop}%
\bibitem [{\citenamefont {Lama}\ \emph {et~al.}(2023)\citenamefont {Lama},
  \citenamefont {Yamamoto}, \citenamefont {Furuta}, \citenamefont {Shimaya},\
  and\ \citenamefont {Takeuchi}}]{Lama_2023}%
  \BibitemOpen
  \bibfield  {author} {\bibinfo {author} {\bibfnamefont {H.}~\bibnamefont
  {Lama}}, \bibinfo {author} {\bibfnamefont {M.~J.}\ \bibnamefont {Yamamoto}},
  \bibinfo {author} {\bibfnamefont {Y.}~\bibnamefont {Furuta}}, \bibinfo
  {author} {\bibfnamefont {T.}~\bibnamefont {Shimaya}},\ and\ \bibinfo {author}
  {\bibfnamefont {K.~A.}\ \bibnamefont {Takeuchi}},\ }\href@noop {} {\bibinfo
  {title} {Emergence of bacterial glass}} (\bibinfo {year} {2023}),\ \Eprint
  {https://arxiv.org/abs/2205.10436} {arXiv:2205.10436 [cond-mat.stat-mech]}
  \BibitemShut {NoStop}%
\bibitem [{\citenamefont {Sugino}\ \emph {et~al.}(2024)\citenamefont {Sugino},
  \citenamefont {Ebata}, \citenamefont {Sowa}, \citenamefont {Ikeda},\ and\
  \citenamefont {Mizuno}}]{Sugino2024}%
  \BibitemOpen
  \bibfield  {author} {\bibinfo {author} {\bibfnamefont {Y.}~\bibnamefont
  {Sugino}}, \bibinfo {author} {\bibfnamefont {H.}~\bibnamefont {Ebata}},
  \bibinfo {author} {\bibfnamefont {Y.}~\bibnamefont {Sowa}}, \bibinfo {author}
  {\bibfnamefont {A.}~\bibnamefont {Ikeda}},\ and\ \bibinfo {author}
  {\bibfnamefont {D.}~\bibnamefont {Mizuno}},\ }\href@noop {} {\bibinfo {title}
  {Non-equilibrium fluidization of dense active suspension}} (\bibinfo {year}
  {2024}),\ \Eprint {https://arxiv.org/abs/arXiv:2401.15658} {arXiv:2401.15658}
  \BibitemShut {NoStop}%
\bibitem [{\citenamefont {Henkes}\ \emph {et~al.}(2011)\citenamefont {Henkes},
  \citenamefont {Fily},\ and\ \citenamefont {Marchetti}}]{Henkes_2011}%
  \BibitemOpen
  \bibfield  {author} {\bibinfo {author} {\bibfnamefont {S.}~\bibnamefont
  {Henkes}}, \bibinfo {author} {\bibfnamefont {Y.}~\bibnamefont {Fily}},\ and\
  \bibinfo {author} {\bibfnamefont {M.~C.}\ \bibnamefont {Marchetti}},\
  }\href@noop {} {\bibfield  {journal} {\bibinfo  {journal} {Physical Review
  E}\ }\textbf {\bibinfo {volume} {84}},\ \bibinfo {pages} {040301} (\bibinfo
  {year} {2011})}\BibitemShut {NoStop}%
\bibitem [{\citenamefont {Bi}\ \emph {et~al.}(2016)\citenamefont {Bi},
  \citenamefont {Yang}, \citenamefont {Marchetti},\ and\ \citenamefont
  {Manning}}]{Bi2016}%
  \BibitemOpen
  \bibfield  {author} {\bibinfo {author} {\bibfnamefont {D.}~\bibnamefont
  {Bi}}, \bibinfo {author} {\bibfnamefont {X.}~\bibnamefont {Yang}}, \bibinfo
  {author} {\bibfnamefont {M.~C.}\ \bibnamefont {Marchetti}},\ and\ \bibinfo
  {author} {\bibfnamefont {M.~L.}\ \bibnamefont {Manning}},\ }\href
  {https://doi.org/10.1103/PhysRevX.6.021011} {\bibfield  {journal} {\bibinfo
  {journal} {Phys. Rev. X}\ }\textbf {\bibinfo {volume} {6}},\ \bibinfo {pages}
  {021011} (\bibinfo {year} {2016})}\BibitemShut {NoStop}%
\bibitem [{\citenamefont {Oyama}\ \emph {et~al.}(2019)\citenamefont {Oyama},
  \citenamefont {Kawasaki}, \citenamefont {Mizuno},\ and\ \citenamefont
  {Ikeda}}]{Oyama2019}%
  \BibitemOpen
  \bibfield  {author} {\bibinfo {author} {\bibfnamefont {N.}~\bibnamefont
  {Oyama}}, \bibinfo {author} {\bibfnamefont {T.}~\bibnamefont {Kawasaki}},
  \bibinfo {author} {\bibfnamefont {H.}~\bibnamefont {Mizuno}},\ and\ \bibinfo
  {author} {\bibfnamefont {A.}~\bibnamefont {Ikeda}},\ }\href
  {https://doi.org/10.1103/PhysRevResearch.1.032038} {\bibfield  {journal}
  {\bibinfo  {journal} {Phys. Rev. Res.}\ }\textbf {\bibinfo {volume} {1}},\
  \bibinfo {pages} {032038} (\bibinfo {year} {2019})}\BibitemShut {NoStop}%
\bibitem [{\citenamefont {Ni}\ \emph {et~al.}(2013)\citenamefont {Ni},
  \citenamefont {Stuart},\ and\ \citenamefont {Dijkstra}}]{Ni2013}%
  \BibitemOpen
  \bibfield  {author} {\bibinfo {author} {\bibfnamefont {R.}~\bibnamefont
  {Ni}}, \bibinfo {author} {\bibfnamefont {M.~A.~C.}\ \bibnamefont {Stuart}},\
  and\ \bibinfo {author} {\bibfnamefont {M.}~\bibnamefont {Dijkstra}},\ }\href
  {https://doi.org/10.1038/ncomms3704} {\bibfield  {journal} {\bibinfo
  {journal} {Nature Communications}\ }\textbf {\bibinfo {volume} {4}},\
  \bibinfo {pages} {2704} (\bibinfo {year} {2013})}\BibitemShut {NoStop}%
\bibitem [{\citenamefont {Berthier}(2014)}]{Berthier_2014}%
  \BibitemOpen
  \bibfield  {author} {\bibinfo {author} {\bibfnamefont {L.}~\bibnamefont
  {Berthier}},\ }\href@noop {} {\bibfield  {journal} {\bibinfo  {journal}
  {Physical review letters}\ }\textbf {\bibinfo {volume} {112}},\ \bibinfo
  {pages} {220602} (\bibinfo {year} {2014})}\BibitemShut {NoStop}%
\bibitem [{\citenamefont {Flenner}\ \emph {et~al.}(2016)\citenamefont
  {Flenner}, \citenamefont {Szamel},\ and\ \citenamefont
  {Berthier}}]{Flenner_2016}%
  \BibitemOpen
  \bibfield  {author} {\bibinfo {author} {\bibfnamefont {E.}~\bibnamefont
  {Flenner}}, \bibinfo {author} {\bibfnamefont {G.}~\bibnamefont {Szamel}},\
  and\ \bibinfo {author} {\bibfnamefont {L.}~\bibnamefont {Berthier}},\
  }\href@noop {} {\bibfield  {journal} {\bibinfo  {journal} {Soft matter}\
  }\textbf {\bibinfo {volume} {12}},\ \bibinfo {pages} {7136} (\bibinfo {year}
  {2016})}\BibitemShut {NoStop}%
\bibitem [{\citenamefont {Nandi}\ and\ \citenamefont {Gov}(2017)}]{Nandi_2017}%
  \BibitemOpen
  \bibfield  {author} {\bibinfo {author} {\bibfnamefont {S.~K.}\ \bibnamefont
  {Nandi}}\ and\ \bibinfo {author} {\bibfnamefont {N.~S.}\ \bibnamefont
  {Gov}},\ }\href@noop {} {\bibfield  {journal} {\bibinfo  {journal} {Soft
  matter}\ }\textbf {\bibinfo {volume} {13}},\ \bibinfo {pages} {7609}
  (\bibinfo {year} {2017})}\BibitemShut {NoStop}%
\bibitem [{\citenamefont {Nandi}\ \emph {et~al.}(2018)\citenamefont {Nandi},
  \citenamefont {Mandal}, \citenamefont {Bhuyan}, \citenamefont {Dasgupta},
  \citenamefont {Rao},\ and\ \citenamefont {Gov}}]{Nandi_2018}%
  \BibitemOpen
  \bibfield  {author} {\bibinfo {author} {\bibfnamefont {S.~K.}\ \bibnamefont
  {Nandi}}, \bibinfo {author} {\bibfnamefont {R.}~\bibnamefont {Mandal}},
  \bibinfo {author} {\bibfnamefont {P.~J.}\ \bibnamefont {Bhuyan}}, \bibinfo
  {author} {\bibfnamefont {C.}~\bibnamefont {Dasgupta}}, \bibinfo {author}
  {\bibfnamefont {M.}~\bibnamefont {Rao}},\ and\ \bibinfo {author}
  {\bibfnamefont {N.~S.}\ \bibnamefont {Gov}},\ }\href@noop {} {\bibfield
  {journal} {\bibinfo  {journal} {Proceedings of the National Academy of
  Sciences}\ }\textbf {\bibinfo {volume} {115}},\ \bibinfo {pages} {7688}
  (\bibinfo {year} {2018})}\BibitemShut {NoStop}%
\bibitem [{\citenamefont {Mandal}\ and\ \citenamefont
  {Sollich}(2020)}]{Mandal_2020}%
  \BibitemOpen
  \bibfield  {author} {\bibinfo {author} {\bibfnamefont {R.}~\bibnamefont
  {Mandal}}\ and\ \bibinfo {author} {\bibfnamefont {P.}~\bibnamefont
  {Sollich}},\ }\href@noop {} {\bibfield  {journal} {\bibinfo  {journal}
  {Physical Review Letters}\ }\textbf {\bibinfo {volume} {125}},\ \bibinfo
  {pages} {218001} (\bibinfo {year} {2020})}\BibitemShut {NoStop}%
\bibitem [{\citenamefont {Klongvessa}\ \emph {et~al.}(2022)\citenamefont
  {Klongvessa}, \citenamefont {Ybert}, \citenamefont {Cottin-Bizonne},
  \citenamefont {Kawasaki},\ and\ \citenamefont {Leocmach}}]{Klongvessa_2022}%
  \BibitemOpen
  \bibfield  {author} {\bibinfo {author} {\bibfnamefont {N.}~\bibnamefont
  {Klongvessa}}, \bibinfo {author} {\bibfnamefont {C.}~\bibnamefont {Ybert}},
  \bibinfo {author} {\bibfnamefont {C.}~\bibnamefont {Cottin-Bizonne}},
  \bibinfo {author} {\bibfnamefont {T.}~\bibnamefont {Kawasaki}},\ and\
  \bibinfo {author} {\bibfnamefont {M.}~\bibnamefont {Leocmach}},\ }\href@noop
  {} {\bibfield  {journal} {\bibinfo  {journal} {The Journal of Chemical
  Physics}\ }\textbf {\bibinfo {volume} {156}} (\bibinfo {year}
  {2022})}\BibitemShut {NoStop}%
\bibitem [{\citenamefont {Paul}\ \emph {et~al.}(2023)\citenamefont {Paul},
  \citenamefont {Mutneja}, \citenamefont {Nandi},\ and\ \citenamefont
  {Karmakar}}]{Paul_2023}%
  \BibitemOpen
  \bibfield  {author} {\bibinfo {author} {\bibfnamefont {K.}~\bibnamefont
  {Paul}}, \bibinfo {author} {\bibfnamefont {A.}~\bibnamefont {Mutneja}},
  \bibinfo {author} {\bibfnamefont {S.~K.}\ \bibnamefont {Nandi}},\ and\
  \bibinfo {author} {\bibfnamefont {S.}~\bibnamefont {Karmakar}},\ }\href@noop
  {} {\bibfield  {journal} {\bibinfo  {journal} {Proceedings of the National
  Academy of Sciences}\ }\textbf {\bibinfo {volume} {120}},\ \bibinfo {pages}
  {e2217073120} (\bibinfo {year} {2023})}\BibitemShut {NoStop}%
\bibitem [{\citenamefont {Keta}\ \emph {et~al.}(2023)\citenamefont {Keta},
  \citenamefont {Mandal}, \citenamefont {Sollich}, \citenamefont {Jack},\ and\
  \citenamefont {Berthier}}]{Keta_2023}%
  \BibitemOpen
  \bibfield  {author} {\bibinfo {author} {\bibfnamefont {Y.-E.}\ \bibnamefont
  {Keta}}, \bibinfo {author} {\bibfnamefont {R.}~\bibnamefont {Mandal}},
  \bibinfo {author} {\bibfnamefont {P.}~\bibnamefont {Sollich}}, \bibinfo
  {author} {\bibfnamefont {R.~L.}\ \bibnamefont {Jack}},\ and\ \bibinfo
  {author} {\bibfnamefont {L.}~\bibnamefont {Berthier}},\ }\href@noop {}
  {\bibfield  {journal} {\bibinfo  {journal} {Soft Matter}\ }\textbf {\bibinfo
  {volume} {19}},\ \bibinfo {pages} {3871} (\bibinfo {year}
  {2023})}\BibitemShut {NoStop}%
\bibitem [{\citenamefont {Paoluzzi}\ \emph {et~al.}(2024)\citenamefont
  {Paoluzzi}, \citenamefont {Levis},\ and\ \citenamefont
  {Pagonabarraga}}]{Paoluzzi2024}%
  \BibitemOpen
  \bibfield  {author} {\bibinfo {author} {\bibfnamefont {M.}~\bibnamefont
  {Paoluzzi}}, \bibinfo {author} {\bibfnamefont {D.}~\bibnamefont {Levis}},\
  and\ \bibinfo {author} {\bibfnamefont {I.}~\bibnamefont {Pagonabarraga}},\
  }\href@noop {} {\bibfield  {journal} {\bibinfo  {journal} {Communications
  Physics}\ }\textbf {\bibinfo {volume} {7}},\ \bibinfo {pages} {57} (\bibinfo
  {year} {2024})}\BibitemShut {NoStop}%
\bibitem [{\citenamefont {Berthier}\ and\ \citenamefont
  {Kurchan}(2013)}]{Berthier_2013}%
  \BibitemOpen
  \bibfield  {author} {\bibinfo {author} {\bibfnamefont {L.}~\bibnamefont
  {Berthier}}\ and\ \bibinfo {author} {\bibfnamefont {J.}~\bibnamefont
  {Kurchan}},\ }\href {https://doi.org/10.1038/nphys2592} {\bibfield  {journal}
  {\bibinfo  {journal} {Nature Physics}\ }\textbf {\bibinfo {volume} {9}},\
  \bibinfo {pages} {310} (\bibinfo {year} {2013})}\BibitemShut {NoStop}%
\bibitem [{\citenamefont {Liluashvili}\ \emph {et~al.}(2017)\citenamefont
  {Liluashvili}, \citenamefont {{\'O}nody},\ and\ \citenamefont
  {Voigtmann}}]{Liluashvili_2017}%
  \BibitemOpen
  \bibfield  {author} {\bibinfo {author} {\bibfnamefont {A.}~\bibnamefont
  {Liluashvili}}, \bibinfo {author} {\bibfnamefont {J.}~\bibnamefont
  {{\'O}nody}},\ and\ \bibinfo {author} {\bibfnamefont {T.}~\bibnamefont
  {Voigtmann}},\ }\href@noop {} {\bibfield  {journal} {\bibinfo  {journal}
  {Physical Review E}\ }\textbf {\bibinfo {volume} {96}},\ \bibinfo {pages}
  {062608} (\bibinfo {year} {2017})}\BibitemShut {NoStop}%
\bibitem [{\citenamefont {Buchenau}\ \emph {et~al.}(1984)\citenamefont
  {Buchenau}, \citenamefont {N$\ddot{\text{u}}$cker},\ and\ \citenamefont
  {Dianoux}}]{Buchenau_1984}%
  \BibitemOpen
  \bibfield  {author} {\bibinfo {author} {\bibfnamefont {U.}~\bibnamefont
  {Buchenau}}, \bibinfo {author} {\bibfnamefont {N.}~\bibnamefont
  {N$\ddot{\text{u}}$cker}},\ and\ \bibinfo {author} {\bibfnamefont {A.~J.}\
  \bibnamefont {Dianoux}},\ }\href@noop {} {\bibfield  {journal} {\bibinfo
  {journal} {Phys. Rev. Lett.}\ }\textbf {\bibinfo {volume} {53}},\ \bibinfo
  {pages} {2316} (\bibinfo {year} {1984})}\BibitemShut {NoStop}%
\bibitem [{\citenamefont {Yamamuro}\ \emph {et~al.}(1996)\citenamefont
  {Yamamuro}, \citenamefont {Matsuo}, \citenamefont {Takeda}, \citenamefont
  {Kanaya}, \citenamefont {Kawaguchi},\ and\ \citenamefont
  {Kaji}}]{Yamamuro_1996}%
  \BibitemOpen
  \bibfield  {author} {\bibinfo {author} {\bibfnamefont {O.}~\bibnamefont
  {Yamamuro}}, \bibinfo {author} {\bibfnamefont {T.}~\bibnamefont {Matsuo}},
  \bibinfo {author} {\bibfnamefont {K.}~\bibnamefont {Takeda}}, \bibinfo
  {author} {\bibfnamefont {T.}~\bibnamefont {Kanaya}}, \bibinfo {author}
  {\bibfnamefont {T.}~\bibnamefont {Kawaguchi}},\ and\ \bibinfo {author}
  {\bibfnamefont {K.}~\bibnamefont {Kaji}},\ }\href
  {https://doi.org/10.1063/1.471928} {\bibfield  {journal} {\bibinfo  {journal}
  {The Journal of Chemical Physics}\ }\textbf {\bibinfo {volume} {105}},\
  \bibinfo {pages} {732} (\bibinfo {year} {1996})}\BibitemShut {NoStop}%
\bibitem [{\citenamefont {Mori}\ \emph {et~al.}(2020)\citenamefont {Mori},
  \citenamefont {Jiang}, \citenamefont {Fujii}, \citenamefont {Kitani},
  \citenamefont {Mizuno}, \citenamefont {Koreeda}, \citenamefont {Motoji},
  \citenamefont {Tokoro}, \citenamefont {Shiraki}, \citenamefont {Yamamoto},\
  and\ \citenamefont {Kojima}}]{Mori_2020}%
  \BibitemOpen
  \bibfield  {author} {\bibinfo {author} {\bibfnamefont {T.}~\bibnamefont
  {Mori}}, \bibinfo {author} {\bibfnamefont {Y.}~\bibnamefont {Jiang}},
  \bibinfo {author} {\bibfnamefont {Y.}~\bibnamefont {Fujii}}, \bibinfo
  {author} {\bibfnamefont {S.}~\bibnamefont {Kitani}}, \bibinfo {author}
  {\bibfnamefont {H.}~\bibnamefont {Mizuno}}, \bibinfo {author} {\bibfnamefont
  {A.}~\bibnamefont {Koreeda}}, \bibinfo {author} {\bibfnamefont
  {L.}~\bibnamefont {Motoji}}, \bibinfo {author} {\bibfnamefont
  {H.}~\bibnamefont {Tokoro}}, \bibinfo {author} {\bibfnamefont
  {K.}~\bibnamefont {Shiraki}}, \bibinfo {author} {\bibfnamefont
  {Y.}~\bibnamefont {Yamamoto}},\ and\ \bibinfo {author} {\bibfnamefont
  {S.}~\bibnamefont {Kojima}},\ }\href
  {https://doi.org/10.1103/PhysRevE.102.022502} {\bibfield  {journal} {\bibinfo
   {journal} {Phys. Rev. E}\ }\textbf {\bibinfo {volume} {102}},\ \bibinfo
  {pages} {022502} (\bibinfo {year} {2020})}\BibitemShut {NoStop}%
\bibitem [{\citenamefont {Lerner}\ \emph {et~al.}(2016)\citenamefont {Lerner},
  \citenamefont {D\"uring},\ and\ \citenamefont {Bouchbinder}}]{Lerner_2016}%
  \BibitemOpen
  \bibfield  {author} {\bibinfo {author} {\bibfnamefont {E.}~\bibnamefont
  {Lerner}}, \bibinfo {author} {\bibfnamefont {G.}~\bibnamefont {D\"uring}},\
  and\ \bibinfo {author} {\bibfnamefont {E.}~\bibnamefont {Bouchbinder}},\
  }\href {https://doi.org/10.1103/PhysRevLett.117.035501} {\bibfield  {journal}
  {\bibinfo  {journal} {Phys. Rev. Lett.}\ }\textbf {\bibinfo {volume} {117}},\
  \bibinfo {pages} {035501} (\bibinfo {year} {2016})}\BibitemShut {NoStop}%
\bibitem [{\citenamefont {Mizuno}\ \emph {et~al.}(2017)\citenamefont {Mizuno},
  \citenamefont {Shiba},\ and\ \citenamefont {Ikeda}}]{Mizuno_2017}%
  \BibitemOpen
  \bibfield  {author} {\bibinfo {author} {\bibfnamefont {H.}~\bibnamefont
  {Mizuno}}, \bibinfo {author} {\bibfnamefont {H.}~\bibnamefont {Shiba}},\ and\
  \bibinfo {author} {\bibfnamefont {A.}~\bibnamefont {Ikeda}},\ }\href
  {https://doi.org/10.1073/pnas.1709015114} {\bibfield  {journal} {\bibinfo
  {journal} {Proceedings of the National Academy of Sciences}\ }\textbf
  {\bibinfo {volume} {114}},\ \bibinfo {pages} {E9767} (\bibinfo {year}
  {2017})}\BibitemShut {NoStop}%
\bibitem [{\citenamefont {Shimada}\ \emph {et~al.}(2018)\citenamefont
  {Shimada}, \citenamefont {Mizuno},\ and\ \citenamefont
  {Ikeda}}]{Shimada_2018}%
  \BibitemOpen
  \bibfield  {author} {\bibinfo {author} {\bibfnamefont {M.}~\bibnamefont
  {Shimada}}, \bibinfo {author} {\bibfnamefont {H.}~\bibnamefont {Mizuno}},\
  and\ \bibinfo {author} {\bibfnamefont {A.}~\bibnamefont {Ikeda}},\ }\href
  {https://doi.org/10.1103/PhysRevE.97.022609} {\bibfield  {journal} {\bibinfo
  {journal} {Phys. Rev. E}\ }\textbf {\bibinfo {volume} {97}},\ \bibinfo
  {pages} {022609} (\bibinfo {year} {2018})}\BibitemShut {NoStop}%
\bibitem [{\citenamefont {Wang}\ \emph {et~al.}(2019)\citenamefont {Wang},
  \citenamefont {Ninarello}, \citenamefont {Guan}, \citenamefont {Berthier},
  \citenamefont {Szamel},\ and\ \citenamefont {Flenner}}]{Wang_2019}%
  \BibitemOpen
  \bibfield  {author} {\bibinfo {author} {\bibfnamefont {L.}~\bibnamefont
  {Wang}}, \bibinfo {author} {\bibfnamefont {A.}~\bibnamefont {Ninarello}},
  \bibinfo {author} {\bibfnamefont {P.}~\bibnamefont {Guan}}, \bibinfo {author}
  {\bibfnamefont {L.}~\bibnamefont {Berthier}}, \bibinfo {author}
  {\bibfnamefont {G.}~\bibnamefont {Szamel}},\ and\ \bibinfo {author}
  {\bibfnamefont {E.}~\bibnamefont {Flenner}},\ }\href@noop {} {\bibfield
  {journal} {\bibinfo  {journal} {Nature Communications}\ }\textbf {\bibinfo
  {volume} {10}},\ \bibinfo {pages} {26} (\bibinfo {year} {2019})}\BibitemShut
  {NoStop}%
\bibitem [{\citenamefont {Mizuno}\ \emph {et~al.}(2020)\citenamefont {Mizuno},
  \citenamefont {Tong}, \citenamefont {Ikeda},\ and\ \citenamefont
  {Mossa}}]{Mizuno_2020}%
  \BibitemOpen
  \bibfield  {author} {\bibinfo {author} {\bibfnamefont {H.}~\bibnamefont
  {Mizuno}}, \bibinfo {author} {\bibfnamefont {H.}~\bibnamefont {Tong}},
  \bibinfo {author} {\bibfnamefont {A.}~\bibnamefont {Ikeda}},\ and\ \bibinfo
  {author} {\bibfnamefont {S.}~\bibnamefont {Mossa}},\ }\href
  {https://doi.org/10.1063/5.0021228} {\bibfield  {journal} {\bibinfo
  {journal} {The Journal of Chemical Physics}\ }\textbf {\bibinfo {volume}
  {153}},\ \bibinfo {pages} {154501} (\bibinfo {year} {2020})}\BibitemShut
  {NoStop}%
\bibitem [{\citenamefont {Widmer-Cooper}\ \emph {et~al.}(2008)\citenamefont
  {Widmer-Cooper}, \citenamefont {Perry}, \citenamefont {Harrowell},\ and\
  \citenamefont {Reichman}}]{Widmer_2008}%
  \BibitemOpen
  \bibfield  {author} {\bibinfo {author} {\bibfnamefont {A.}~\bibnamefont
  {Widmer-Cooper}}, \bibinfo {author} {\bibfnamefont {H.}~\bibnamefont
  {Perry}}, \bibinfo {author} {\bibfnamefont {P.}~\bibnamefont {Harrowell}},\
  and\ \bibinfo {author} {\bibfnamefont {D.~R.}\ \bibnamefont {Reichman}},\
  }\href@noop {} {\bibfield  {journal} {\bibinfo  {journal} {Nature phys.}\
  }\textbf {\bibinfo {volume} {4}},\ \bibinfo {pages} {711} (\bibinfo {year}
  {2008})}\BibitemShut {NoStop}%
\bibitem [{\citenamefont {Martin}\ \emph {et~al.}(2021)\citenamefont {Martin},
  \citenamefont {O'Byrne}, \citenamefont {Cates}, \citenamefont {Fodor},
  \citenamefont {Nardini}, \citenamefont {Tailleur},\ and\ \citenamefont {van
  Wijland}}]{Martin_2021}%
  \BibitemOpen
  \bibfield  {author} {\bibinfo {author} {\bibfnamefont {D.}~\bibnamefont
  {Martin}}, \bibinfo {author} {\bibfnamefont {J.}~\bibnamefont {O'Byrne}},
  \bibinfo {author} {\bibfnamefont {M.~E.}\ \bibnamefont {Cates}}, \bibinfo
  {author} {\bibfnamefont {E.}~\bibnamefont {Fodor}}, \bibinfo {author}
  {\bibfnamefont {C.}~\bibnamefont {Nardini}}, \bibinfo {author} {\bibfnamefont
  {J.}~\bibnamefont {Tailleur}},\ and\ \bibinfo {author} {\bibfnamefont
  {F.}~\bibnamefont {van Wijland}},\ }\href
  {https://doi.org/10.1103/PhysRevE.103.032607} {\bibfield  {journal} {\bibinfo
   {journal} {Phys. Rev. E}\ }\textbf {\bibinfo {volume} {103}},\ \bibinfo
  {pages} {032607} (\bibinfo {year} {2021})}\BibitemShut {NoStop}%
\bibitem [{\citenamefont {Fily}\ and\ \citenamefont
  {Marchetti}(2012)}]{Fily2012}%
  \BibitemOpen
  \bibfield  {author} {\bibinfo {author} {\bibfnamefont {Y.}~\bibnamefont
  {Fily}}\ and\ \bibinfo {author} {\bibfnamefont {M.~C.}\ \bibnamefont
  {Marchetti}},\ }\href@noop {} {\bibfield  {journal} {\bibinfo  {journal}
  {Physical review letters}\ }\textbf {\bibinfo {volume} {108}},\ \bibinfo
  {pages} {235702} (\bibinfo {year} {2012})}\BibitemShut {NoStop}%
\bibitem [{\citenamefont {Solon}\ \emph {et~al.}(2015)\citenamefont {Solon},
  \citenamefont {Cates},\ and\ \citenamefont {Tailleur}}]{Solon_2015}%
  \BibitemOpen
  \bibfield  {author} {\bibinfo {author} {\bibfnamefont {A.~P.}\ \bibnamefont
  {Solon}}, \bibinfo {author} {\bibfnamefont {M.~E.}\ \bibnamefont {Cates}},\
  and\ \bibinfo {author} {\bibfnamefont {J.}~\bibnamefont {Tailleur}},\
  }\href@noop {} {\bibfield  {journal} {\bibinfo  {journal} {The European
  Physical Journal Special Topics}\ }\textbf {\bibinfo {volume} {224}},\
  \bibinfo {pages} {1231} (\bibinfo {year} {2015})}\BibitemShut {NoStop}%
\bibitem [{\citenamefont {Henkes}\ \emph {et~al.}(2020)\citenamefont {Henkes},
  \citenamefont {Kostanjevec}, \citenamefont {Collinson}, \citenamefont
  {Sknepnek},\ and\ \citenamefont {Bertin}}]{Henkes_2020}%
  \BibitemOpen
  \bibfield  {author} {\bibinfo {author} {\bibfnamefont {S.}~\bibnamefont
  {Henkes}}, \bibinfo {author} {\bibfnamefont {K.}~\bibnamefont {Kostanjevec}},
  \bibinfo {author} {\bibfnamefont {J.~M.}\ \bibnamefont {Collinson}}, \bibinfo
  {author} {\bibfnamefont {R.}~\bibnamefont {Sknepnek}},\ and\ \bibinfo
  {author} {\bibfnamefont {E.}~\bibnamefont {Bertin}},\ }\href@noop {}
  {\bibfield  {journal} {\bibinfo  {journal} {Nature communications}\ }\textbf
  {\bibinfo {volume} {11}},\ \bibinfo {pages} {1405} (\bibinfo {year}
  {2020})}\BibitemShut {NoStop}%
\bibitem [{\citenamefont {Ikeda}\ and\ \citenamefont
  {Berthier}(2015)}]{Ikeda_2015}%
  \BibitemOpen
  \bibfield  {author} {\bibinfo {author} {\bibfnamefont {A.}~\bibnamefont
  {Ikeda}}\ and\ \bibinfo {author} {\bibfnamefont {L.}~\bibnamefont
  {Berthier}},\ }\href {https://doi.org/10.1103/PhysRevE.92.012309} {\bibfield
  {journal} {\bibinfo  {journal} {Phys. Rev. E}\ }\textbf {\bibinfo {volume}
  {92}},\ \bibinfo {pages} {012309} (\bibinfo {year} {2015})}\BibitemShut
  {NoStop}%
\bibitem [{\citenamefont {Gardiner}(2009)}]{Gardiner2009}%
  \BibitemOpen
  \bibfield  {author} {\bibinfo {author} {\bibfnamefont {C.}~\bibnamefont
  {Gardiner}},\ }\href@noop {} {\emph {\bibinfo {title} {Stochastic
  methods}}},\ Vol.~\bibinfo {volume} {4}\ (\bibinfo  {publisher} {Springer
  Berlin},\ \bibinfo {year} {2009})\BibitemShut {NoStop}%
\bibitem [{\citenamefont {Santra}\ \emph {et~al.}(2020)\citenamefont {Santra},
  \citenamefont {Basu},\ and\ \citenamefont {Sabhapandit}}]{Santra_2020}%
  \BibitemOpen
  \bibfield  {author} {\bibinfo {author} {\bibfnamefont {I.}~\bibnamefont
  {Santra}}, \bibinfo {author} {\bibfnamefont {U.}~\bibnamefont {Basu}},\ and\
  \bibinfo {author} {\bibfnamefont {S.}~\bibnamefont {Sabhapandit}},\
  }\bibfield  {journal} {\bibinfo  {journal} {Physical Review E}\ }\textbf
  {\bibinfo {volume} {101}},\ \href
  {https://doi.org/10.1103/physreve.101.062120} {10.1103/physreve.101.062120}
  (\bibinfo {year} {2020})\BibitemShut {NoStop}%
\bibitem [{\citenamefont {Bitzek}\ \emph {et~al.}(2006)\citenamefont {Bitzek},
  \citenamefont {Koskinen}, \citenamefont {G\"ahler}, \citenamefont {Moseler},\
  and\ \citenamefont {Gumbsch}}]{Bitzek_2006}%
  \BibitemOpen
  \bibfield  {author} {\bibinfo {author} {\bibfnamefont {E.}~\bibnamefont
  {Bitzek}}, \bibinfo {author} {\bibfnamefont {P.}~\bibnamefont {Koskinen}},
  \bibinfo {author} {\bibfnamefont {F.}~\bibnamefont {G\"ahler}}, \bibinfo
  {author} {\bibfnamefont {M.}~\bibnamefont {Moseler}},\ and\ \bibinfo {author}
  {\bibfnamefont {P.}~\bibnamefont {Gumbsch}},\ }\href
  {https://doi.org/10.1103/PhysRevLett.97.170201} {\bibfield  {journal}
  {\bibinfo  {journal} {Phys. Rev. Lett.}\ }\textbf {\bibinfo {volume} {97}},\
  \bibinfo {pages} {170201} (\bibinfo {year} {2006})}\BibitemShut {NoStop}%
\bibitem [{\citenamefont {Mizuno}\ and\ \citenamefont
  {Ikeda}(2022)}]{MizunoIkeda2022}%
  \BibitemOpen
  \bibfield  {author} {\bibinfo {author} {\bibfnamefont {H.}~\bibnamefont
  {Mizuno}}\ and\ \bibinfo {author} {\bibfnamefont {A.}~\bibnamefont {Ikeda}},\
  }\bibinfo {title} {{Computational Simulations of the Vibrational Properties
  of Glasses}},\ in\ \href {https://doi.org/10.1142/9781800612587_0010} {\emph
  {\bibinfo {booktitle} {{Low-Temperature Thermal and Vibrational Properties of
  Disordered Solids}}}},\ \bibinfo {editor} {edited by\ \bibinfo {editor}
  {\bibfnamefont {M.~A.}\ \bibnamefont {Ramos}}}\ (\bibinfo  {publisher}
  {{WORLD SCIENTIFIC (EUROPE)}},\ \bibinfo {year} {2022})\ Chap.~\bibinfo
  {chapter} {10}, pp.\ \bibinfo {pages} {375--433}\BibitemShut {NoStop}%
\bibitem [{\citenamefont {Gillespie}(1996)}]{Gillespie_1996}%
  \BibitemOpen
  \bibfield  {author} {\bibinfo {author} {\bibfnamefont {D.~T.}\ \bibnamefont
  {Gillespie}},\ }\href {https://doi.org/10.1103/PhysRevE.54.2084} {\bibfield
  {journal} {\bibinfo  {journal} {Phys. Rev. E}\ }\textbf {\bibinfo {volume}
  {54}},\ \bibinfo {pages} {2084} (\bibinfo {year} {1996})}\BibitemShut
  {NoStop}%
\bibitem [{\citenamefont {Coslovich}\ and\ \citenamefont
  {Ikeda}(2022)}]{Coslovich2022}%
  \BibitemOpen
  \bibfield  {author} {\bibinfo {author} {\bibfnamefont {D.}~\bibnamefont
  {Coslovich}}\ and\ \bibinfo {author} {\bibfnamefont {A.}~\bibnamefont
  {Ikeda}},\ }\href@noop {} {\bibfield  {journal} {\bibinfo  {journal} {The
  Journal of Chemical Physics}\ }\textbf {\bibinfo {volume} {156}} (\bibinfo
  {year} {2022})}\BibitemShut {NoStop}%
\bibitem [{\citenamefont {Caprini}\ \emph {et~al.}(2023)\citenamefont
  {Caprini}, \citenamefont {Marini Bettolo~Marconi}, \citenamefont {Puglisi},\
  and\ \citenamefont {L{\"o}wen}}]{Caprini2023}%
  \BibitemOpen
  \bibfield  {author} {\bibinfo {author} {\bibfnamefont {L.}~\bibnamefont
  {Caprini}}, \bibinfo {author} {\bibfnamefont {U.}~\bibnamefont {Marini
  Bettolo~Marconi}}, \bibinfo {author} {\bibfnamefont {A.}~\bibnamefont
  {Puglisi}},\ and\ \bibinfo {author} {\bibfnamefont {H.}~\bibnamefont
  {L{\"o}wen}},\ }\href@noop {} {\bibfield  {journal} {\bibinfo  {journal} {The
  Journal of Chemical Physics}\ }\textbf {\bibinfo {volume} {159}} (\bibinfo
  {year} {2023})}\BibitemShut {NoStop}%
\bibitem [{\citenamefont {Lemons}(2003)}]{Lemons_2003}%
  \BibitemOpen
  \bibfield  {author} {\bibinfo {author} {\bibfnamefont {D.~S.}\ \bibnamefont
  {Lemons}},\ }\href@noop {} {\bibinfo {title} {An introduction to stochastic
  processes in physics}} (\bibinfo {year} {2003})\BibitemShut {NoStop}%
\bibitem [{\citenamefont {Malakar}\ \emph {et~al.}(2020)\citenamefont
  {Malakar}, \citenamefont {Das}, \citenamefont {Kundu}, \citenamefont
  {Kumar},\ and\ \citenamefont {Dhar}}]{Malakar_2020}%
  \BibitemOpen
  \bibfield  {author} {\bibinfo {author} {\bibfnamefont {K.}~\bibnamefont
  {Malakar}}, \bibinfo {author} {\bibfnamefont {A.}~\bibnamefont {Das}},
  \bibinfo {author} {\bibfnamefont {A.}~\bibnamefont {Kundu}}, \bibinfo
  {author} {\bibfnamefont {K.~V.}\ \bibnamefont {Kumar}},\ and\ \bibinfo
  {author} {\bibfnamefont {A.}~\bibnamefont {Dhar}},\ }\href@noop {} {\bibfield
   {journal} {\bibinfo  {journal} {Physical Review E}\ }\textbf {\bibinfo
  {volume} {101}},\ \bibinfo {pages} {022610} (\bibinfo {year}
  {2020})}\BibitemShut {NoStop}%
\bibitem [{\citenamefont {Dhar}\ \emph {et~al.}(2019)\citenamefont {Dhar},
  \citenamefont {Kundu}, \citenamefont {Majumdar}, \citenamefont
  {Sabhapandit},\ and\ \citenamefont {Schehr}}]{Dhar_2019}%
  \BibitemOpen
  \bibfield  {author} {\bibinfo {author} {\bibfnamefont {A.}~\bibnamefont
  {Dhar}}, \bibinfo {author} {\bibfnamefont {A.}~\bibnamefont {Kundu}},
  \bibinfo {author} {\bibfnamefont {S.~N.}\ \bibnamefont {Majumdar}}, \bibinfo
  {author} {\bibfnamefont {S.}~\bibnamefont {Sabhapandit}},\ and\ \bibinfo
  {author} {\bibfnamefont {G.}~\bibnamefont {Schehr}},\ }\href@noop {}
  {\bibfield  {journal} {\bibinfo  {journal} {Physical Review E}\ }\textbf
  {\bibinfo {volume} {99}},\ \bibinfo {pages} {032132} (\bibinfo {year}
  {2019})}\BibitemShut {NoStop}%
\bibitem [{\citenamefont {Pusey}\ and\ \citenamefont
  {Van~Megen}(1986)}]{Pusey1986}%
  \BibitemOpen
  \bibfield  {author} {\bibinfo {author} {\bibfnamefont {P.~N.}\ \bibnamefont
  {Pusey}}\ and\ \bibinfo {author} {\bibfnamefont {W.}~\bibnamefont
  {Van~Megen}},\ }\href@noop {} {\bibfield  {journal} {\bibinfo  {journal}
  {Nature}\ }\textbf {\bibinfo {volume} {320}},\ \bibinfo {pages} {340}
  (\bibinfo {year} {1986})}\BibitemShut {NoStop}%
\bibitem [{\citenamefont {Brito}\ and\ \citenamefont
  {Wyart}(2009)}]{Brito2009}%
  \BibitemOpen
  \bibfield  {author} {\bibinfo {author} {\bibfnamefont {C.}~\bibnamefont
  {Brito}}\ and\ \bibinfo {author} {\bibfnamefont {M.}~\bibnamefont {Wyart}},\
  }\href@noop {} {\bibfield  {journal} {\bibinfo  {journal} {The Journal of
  chemical physics}\ }\textbf {\bibinfo {volume} {131}} (\bibinfo {year}
  {2009})}\BibitemShut {NoStop}%
\bibitem [{\citenamefont {Henkes}\ \emph {et~al.}(2012)\citenamefont {Henkes},
  \citenamefont {Brito},\ and\ \citenamefont {Dauchot}}]{Henkes_2012}%
  \BibitemOpen
  \bibfield  {author} {\bibinfo {author} {\bibfnamefont {S.}~\bibnamefont
  {Henkes}}, \bibinfo {author} {\bibfnamefont {C.}~\bibnamefont {Brito}},\ and\
  \bibinfo {author} {\bibfnamefont {O.}~\bibnamefont {Dauchot}},\ }\href@noop
  {} {\bibfield  {journal} {\bibinfo  {journal} {Soft Matter}\ }\textbf
  {\bibinfo {volume} {8}},\ \bibinfo {pages} {6092} (\bibinfo {year}
  {2012})}\BibitemShut {NoStop}%
\end{thebibliography}%

\end{document}